\documentclass[aps,prd,twocolumn,superscriptaddress,showpacs,amsmath,amssymb]{revtex4-2}

\usepackage{physics}
\usepackage{graphicx}
\graphicspath{{./figures/}}
\usepackage[colorlinks=true,allcolors=blue]{hyperref}%

\DeclareMathOperator{\sgn}{sgn}
\renewcommand{\vec}[1]{\boldsymbol{#1}}
\renewcommand{\paragraph}[1]{\emph{#1}}
\newcommand{\sA}{\alpha} 
\newcommand{\sB}{\beta}

\usepackage{tikz}
\tikzset{every picture/.style={line width=0.2mm}}
\definecolor{pred}{RGB}{238,28,37}
\definecolor{pblue}{RGB}{48,49,146}
\definecolor{pgreen}{RGB}{00,163,80}
\usetikzlibrary{decorations.pathmorphing}
\tikzset{snake it/.style={decorate, decoration=snake}}

\begin{document}

\newcommand{\JLU}{Institut f\"ur Theoretische Physik,
  Justus-Liebig-Universit\"at, 
  35392 Giessen, Germany}   
\newcommand{\HFHF}{Helmholtz Research Academy Hesse for FAIR (HFHF), Campus Giessen, 35392 Giessen, Germany}
\newcommand{\UB}{Fakult\"at f\"ur Physik, Universit\"at Bielefeld, D-33615 Bielefeld, Germany}
\newcommand{\FSU}{Theoretisch-Physikalisches Institut, Abbe Center of Photonics, Friedrich-Schiller-Universit\"at Jena, Max-Wien-Platz 1, 07743 Jena, Germany}

\title{Universal scaling of transport coefficients near the liquid-gas critical point}

\author{Johannes V. Roth}
\affiliation{\UB}

\author{Yunxin Ye}
\affiliation{\FSU}

\author{S\"oren Schlichting}
\affiliation{\UB}

\author{Lorenz von Smekal}
\affiliation{\JLU}
\affiliation{\HFHF}

\date{March 2026}

\begin{abstract}

We employ a novel real-time formulation of the functional renormalization group (FRG) to compute universal scaling functions of the thermal diffusivity and the shear viscosity in the vicinity of the liquid-gas critical point,
i.e.,~for the dynamic universality class of Model~H from the Halperin-Hohenberg classification. We map out the universal dependence of the transport coefficients on temperature, external magnetic field, and wavenumber, and provide a detailed comparison with the Kawasaki approximation, which is here obtained from a perturbative one-loop approximation to our real-time FRG flow. 
In contrast to the Kawasaki approximation, the non-perturbative scaling functions from the full real-time FRG flow show a mild dependence on the thermodynamic path towards the critical point.
We further compare our FRG results for the universal wavenumber and temperature dependence of the thermal diffusivity with experimental data from critical fluids. 
\end{abstract}

\maketitle

\paragraph{Introduction.}
Near a critical point not only thermodynamic but also transport properties become universal.
A classic example is $\mathrm{H_2O}$, where universal scaling functions 
are used by the International Association for the Properties of Water and Steam (IAPWS) to describe the thermal diffusivity and the shear viscosity in the region around the liquid-gas critical point \cite{10.1063/1.3088050,10.1063/1.4738955}. Moreover, the notoriously small critical exponent of the shear viscosity was measured to high precision in the delicate Critical Viscosity of Xenon (CVX) experiment aboard the Space Shuttle Mission STS-85 \cite{PhysRevE.60.4079}. In the theory of dynamic critical phenomena, the liquid-gas critical point 
falls into the dynamic universality class of Model~H from  the Halperin-Hohenberg classification \cite{Hohenberg:1977ym}, which describes a conserved order parameter
that is subject to advection due to thermal velocity fluctuations in the surrounding fluid \cite{Onuki_2002}.

For instance, the universal wavenumber dependence of the thermal diffusivity was measured in a variety of fluids \cite{Swinney:1973zz} and has recently become relevant for the ongoing search of the conjectured QCD critical point at finite baryon chemical potential \cite{Stephanov:2017ghc}. 
Potential signatures of the QCD critical point in heavy-ion collisions are remnants of critical scaling \cite{Stephanov:1998dy}. For instance, arguments based on the Ising universality class would suggest that certain cumulants of the net-baryon number such as the kurtosis show a non-monotonic dependence on beam energy \cite{Stephanov:2011pb}, assuming that the fireball is in local equilibrium at freeze out. However, critical slowing down entails that the critical mode is guaranteed to fall out of local equilibrium as the fireball traverses the critical point \cite{Berdnikov:1999ph}. As such, possible signatures of a critical point in heavy-ion collisions could be mainly determined by its \emph{dynamic} universality class, which is plausibly the one of Model~H \cite{Son:2004iv}.

Historically, universal scaling functions of Model~H were based on the Kawasaki approximation \cite{Kawasaki:1970dpc}, which neglects the (weak) power-law divergence of the shear viscosity. The Kawasaki approximation in its original form and its subsequent improvements \cite{PhysRevLett.29.48,10.1143/PTP.55.1384,10.1143/PTP.64.536} were rather successful in describing 
the universal scaling of transport coefficients 
in critical fluids, see, e.g.,~Chapter 6 of Ref.~\cite{Onuki_2002} for a compilation of experimental and theoretical results. 
However, despite its merits,
the Kawasaki approximation is not based on a systematic expansion.
As such, it would be worthwhile to have a non-perturbative framework that can assess the quality of the Kawasaki approximation. 
Based on the success of the functional renormalization group (FRG) in describing static critical phenomena~\cite{Balog:2019rrg}, such a framework is provided by a real-time formulation of the FRG that was developed in Ref.~\cite{Roth:2024rbi} and applied to Model~H in Ref.~\cite{Roth:2024hcu} (for an independent related study, see~\cite{Chen:2024lzz}), which can be regarded as an approach complementary to direct numerical simulations \cite{Florio:2021jlx,Chattopadhyay:2024jlh}.

In this letter, we use our real-time FRG framework to map out scaling functions in Model~H which describe the universal dependence of the thermal diffusivity and the shear viscosity on temperature, external (magnetic) field, and wavenumber. 
We provide detailed comparisons with the Kawasaki approximation \cite{Kawasaki:1970dpc} and experimental data from fluids near the liquid-gas critical point \cite{Swinney:1973zz}. 

\paragraph{Real-time FRG approach to Model H.} In Model H, the conserved one-component order parameter $\phi$ (usually regarded as the entropy per particle) couples reversibly to the transverse component of the conserved momentum density $\vec{j}$. This is expressed by the equations of motion \cite{PhysRevB.13.1299,Hohenberg:1977ym}
\begin{align}
    \frac{\partial \phi}{\partial t}  &= \sigma \vec{\nabla}^2 \frac{\delta F}{\delta \phi} - g\frac{\delta F}{\delta \vec{j}}\cdot\vec{\nabla}\phi + \theta  \, ,\label{eq:eomsH} \\ \nonumber 
    \frac{\partial \vec{j}}{\partial t} &= \mathcal{T} \bigg[ \bar{\eta} \vec{\nabla}^2 \frac{\delta F}{\delta \vec{j}} + g\frac{\delta F}{\delta \phi} \vec{\nabla}\phi -  g\left(\frac{\delta F}{\delta \vec{j}} \cdot \vec{\nabla}\right)\vec{j} \bigg]   +  \vec{\xi} \, ,
\end{align}
with the effective Hamiltonian 
\begin{equation}
    F =\!\int \! d^d x \left\{ \frac{1}{2}(\vec{\nabla} \phi)^2+\frac{m^2}{2} \phi^2 + \frac{\lambda}{4!}\phi^4- H\phi  + \frac{\vec{j}^2}{2w}  \right\} \label{eq:freeEnergy}
\end{equation}
which consists of a Landau-Ginzburg-Wilson (LGW) model for the order parameter $\phi$, an external symmetry-breaking field $H$, and the kinetic energy of the fluid with mass density $w$.
$\mathcal{T}[\ldots]$ is a projector which selects the transverse part of the vector (field) in brackets. 
The kinetic coefficients $\sigma$ and $\bar{\eta}$ are proportional to the heat conductivity and the shear viscosity of the fluid, respectively. $\theta$ and $\vec{\xi}$ are Gaussian noise terms with vanishing expectation values and variances given by
\begin{align}
    \langle \theta(t,\vec{x})\theta(t',\vec{x}')\rangle & =  -2 \sigma T \, \vec{\nabla}^2 \delta(\vec{x}-\vec{x}') \delta(t-t') \, ,\\
    \langle \xi_l(t,\vec{x})\xi_m(t',\vec{x}')\rangle & = -2 \bar{\eta} T \, \mathcal{T}_{lm} \vec{\nabla}^2 \delta(\vec{x}-\vec{x}') \delta(t-t') \,,\nonumber
\end{align}
which are fixed by the fluctuation-dissipation theorem, ensuring that the system approaches the equilibrium distribution $\propto e^{-F/T}$.

The difficulty in solving the equations of motion \eqref{eq:eomsH} near the critical point are the coherent long-distance fluctuations of the order parameter related to the divergent correlation length, which render the system non-perturbative in $d=3$ spatial dimensions. The idea of the functional renormalization group (FRG) is to artifically suppress fluctuations of modes with wavenumbers $p \lesssim k$ smaller than the FRG scale $k$. This is realized by implementing an infrared regulator $R_k^{\phi}(p)$ in
the equilibrium distribution $e^{-F/T}$, replacing $F \to F +\Delta F_k$ with
\begin{align}
    \Delta F_k[\phi] \equiv \frac{1}{2}\int \frac{d^dp}{(2\pi)^d} \phi(-\vec{p}) R_k^{\phi}(p)\phi(\vec{p}) \,. \label{eq:regInFreeEn}
\end{align}
The regulator $R_k^{\phi}(p)$ has the formal properties of being of order $\sim k^2$ for $p \ll k$, and going to zero for $p \gg k$.

In the pioneering work of Wetterich \cite{Wetterich:1992yh}, the FRG flow is formulated for the (Gibbs) free energy $F_k[\phi,\vec{j}]$, which, due to the regulator, is coarse grained over the length scale $\sim k^{-1}$. 
In particular, the effects of critical fluctuations are suppressed, since the $k$-dependent correlation length $\xi_{k}$ is limited to $\xi_{k} \lesssim k^{-1}$. As such, for a sufficiently large initial scale $k=\Lambda$ a saddle-point approximation becomes exact, and we have $F_{\Lambda} = F$. Lowering the FRG scale $k$ has the effect of integrating out (critical) fluctuations from modes with wavevectors $p \sim k$ where the regulator derivative $\partial_k R_k^{\phi}(p)$ is peaked, as expressed by the exact flow equation \cite{Wetterich:1992yh}
\begin{equation}
    \partial_k F_k[\phi] = \frac{T}{2}  \int \!\frac{d^dp}{(2\pi)^d} \partial_k R_k^{\phi}(p) \left(\frac{\delta^2 F_k[\phi]}{\delta \phi\delta \phi}+R_k^{\phi}\right)^{-1}_{-\vec{p},\vec{p}} \,. \label{eq:flowOfFk}
\end{equation}
At $k=0$ all fluctuations are included, such that $F_{k=0}$ corresponds to the full free energy of the system. However, Eq.~\eqref{eq:flowOfFk} really represents an infinite tower of coupled ordinary differential equations, which has to be truncated in practice. An important example of a systematic truncation scheme is a derivative expansion of $F_k$, which has been shown to yield rapidly convergent results for critical exponents and amplitude ratios 
\cite{DePolsi:2020pjk}. In particular, already the lowest order, which is commonly referred to as local potential approximation (LPA), yields semi-quantitative results for the critical exponents of the $3d$ Ising model (see, e.g.,~Ref.~\cite{Murgana:2023xrq} for a recent comparison).

The basic idea of the FRG also holds for the real-time dynamics of the system, with the regulator effectively coarse graining the equations of motion \eqref{eq:eomsH} over distances $\sim k^{-1}$.
In particular, the prescription \eqref{eq:regInFreeEn} of introducing the regulator on the level of the effective Hamiltonian affects the equations of motion \eqref{eq:eomsH} in a non-trivial way, since the regulator appears not only in the diffusion terms, but also in the reversible (Poisson bracket) terms. In the Martin-Siggia-Rose (MSR) path integral reformulation, which is the basis for the real-time FRG approach, this ensures $(i)$ that the Poisson-bracket structure is preserved, and $(ii)$ that the system is in thermal equilibrium at all FRG scales~$k$ \cite{Roth:2024rbi,Roth:2024hcu}.
In particular, $(i)$ implies that the mode coupling $g$ is independent of the FRG scale~$k$, and $(ii)$ implies that the FRG flow of the coarse-grained free energy $F_k$ stays independent of the dynamics, i.e.,~it still satisfies the closed flow equation
\eqref{eq:flowOfFk} \cite{Roth:2024rbi}.

In this work, we are interested in the universal wavevector  dependence of transport coefficients. Hence, we supplement the LPA with a wavenumber ($p$) dependent wave function renormalization $Z_k(p)$,
\begin{align}
    F_k &= \frac{1}{2}\int \frac{d^d p}{(2\pi)^d} \,\phi(-\vec{p}) Z_k(p)\vec{p}^2 \phi(\vec{p}) \label{eq:freeEnergyTrunc} \\ \nonumber
    &\hspace{2cm} + \int d^d x \bigg\{ U_k(\phi^2)  - H\phi  + \frac{\vec{j}^2}{2w}\bigg\} 
\end{align}
where $U_k(\phi^2)$ is the effective potential of the order parameter. As such, our truncation includes the full field-dependent effective potential and a field-independent but wavenumber-dependent wave function renormalization. 
We truncate the dynamic sector by promoting the kinetic coefficients in the equations of motion \eqref{eq:eomsH} to depend on wavenumber $p$. In coordinate space, this amounts to the coarse-grained equations of motion \footnote{Notice that the Langevin noise terms have disappeared since the coarse-grained equations of motion hold for the expectation values of the fields (which we here denote by the same symbols $\phi$ and $\vec{j}$ for simplicity).}
\begin{align}
    \frac{\partial \phi}{\partial t}  &= -\gamma_{\phi,k}(-i\vec{\nabla}) \frac{\delta F_k}{\delta \phi} + g\{\phi,\vec{j}\} \cdot \frac{\delta F_k}{\delta \vec{j}}  \, ,\label{eq:eomsHTrunc} \\ \nonumber 
    \frac{\partial \vec{j}}{\partial t} &= \mathcal{T} \bigg[ -\gamma_{j,k}(-i\vec{\nabla}) \frac{\delta F_k}{\delta \vec{j}} \!+\! g\{\vec{j},\phi\} \frac{\delta F_k}{\delta \phi} \!+\!  g\{\vec{j},j_n\}\frac{\delta F_k}{\delta j_n} \bigg]  \, ,
\end{align}
for the fields' expectation values. 
Since both $\phi$ and $\vec{j}$ are densities of conserved quantities, the Taylor expansions of the Fourier-transformed kinetic coefficients $\gamma_{\phi,k}(\vec{p}) = \sigma_k \vec{p}^2 + \cdots$ and $\gamma_{j,k}(\vec{p}) = \bar{\eta}_k \vec{p}^2 + \cdots$ start at order $\vec{p}^2$.

In practice, we solve the FRG flow numerically by expanding the flow equations 
around the $k$-dependent minimum of the coarse-grained free energy \eqref{eq:freeEnergyTrunc} \cite{SuppMat}. For the remainder of this work, all quantities are understood at $k=0$, unless explicitly stated otherwise.

\paragraph{Static critical behavior.} 
Sufficiently close to the critical point, observables are described by homogeneous functions of the relevant couplings, which here are the reduced temperature $\tau \equiv (T-T_c)/T_c$ and the dimensionless external field $h \equiv H/H_0$. In case of the order parameter $\phi$, this is expressed by Widom's scaling hypothesis,
\begin{align}
    \phi(\tau,h) = s^{-\beta/\nu} \phi(s^{1/\nu}\tau, s^{\beta\delta/\nu}h) \,,
\end{align}
which implies that the magnetic scaling of the order parameter can be described by the Griffiths scaling function 
\begin{align}
    \phi = B^c h^{1/\delta} f_G(z) \label{eq:phiScaling}
\end{align}
with the scaling variable $z \equiv \tau h^{-1/\beta\delta}$. The non-universal amplitude $B^c$ is fixed by requiring $f_G(0) = 1$, which fixes the power-law behavior of the order parameter at $T=T_c$,
\begin{equation}
    \phi = B^c h^{1/\delta} + \cdots \,, \quad H \to 0^+ \,. \label{eq:phiAtTc}
\end{equation}
On the other hand, when the critical temperature is approached at $H=0$ from below, the order parameter behaves as
\begin{equation}
    \phi = B (-\tau)^\beta + \cdots \,, \quad \tau\to 0^- \label{eq:phiAtH0}
\end{equation}
which defines the non-universal amplitude $B$. In practice, we look for a plateau in logarithmic derivatives of \eqref{eq:phiAtTc} and \eqref{eq:phiAtH0}, for which we find $\delta \approx 4.39$ and $\beta \approx 0.358$, respectively \footnote{We employ the initial conditions $m^2_{\Lambda}=-1$, $\lambda_{\Lambda}=1$, $w=1$ at the UV scale $\Lambda=\pi$. Power-law fits in the scaling regime of \protect\eqref{eq:phiAtTc} and \protect\eqref{eq:phiAtH0} yield the non-universal amplitudes $B^c/H_0^{1/\delta} \approx 2.03$ and $B \approx 2.60$. The non-universal amplitude $H_0$ is fixed by the normalization $f_G(z)/(-z)^{\beta} \to 1$ for $z \to -\infty$, which is equivalent to $B=B^c$ and yields $H_0=(B/(B^c/H_0^{1/\delta}))^{\delta}\approx 2.99$.}.

\begin{figure}[t]
    \centering
    \includegraphics[width=\linewidth]{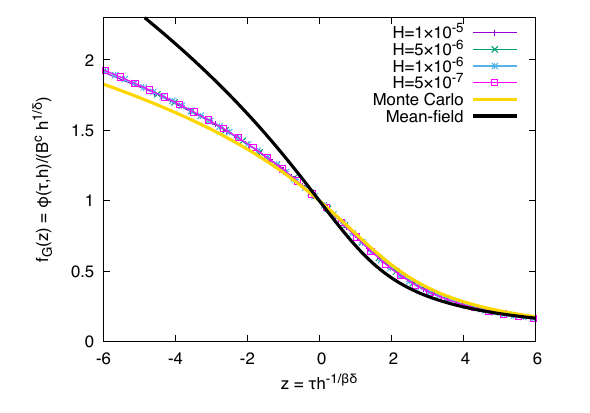}
    \caption{Magnetic scaling function $f_G(z)$ of the order parameter as defined in \eqref{eq:phiScaling}, comparing our rescaled FRG results with the result of Monte Carlo simulations \cite{Karsch:2023pga} and the mean-field scaling function. The normalization is fixed by $f_G(0) = 1$ and $f_G(z)/(-z)^{\beta} \to 1$ for $z \to-\infty$.
    \label{fig:scalingPhi0}}
\end{figure}

To obtain the universal scaling function $f_G(z)$, we compute the expectation value of the order parameter $\phi(\tau,h)$ along various isomagnetic lines ($h=const$) near the critical point in the phase diagram. We rescale the resulting curves according to \eqref{eq:phiScaling}. The results are shown in Fig.~\ref{fig:scalingPhi0}, together with results from Monte Carlo (MC) simulations \cite{Karsch:2023pga}, and the solution to the mean-field (MF) equation $z f_G(z) + f_G^3(z) = 1$. The latter corresponds to the critical exponents $\beta_{\text{MF}}=1/2$ and $\delta_{\text{MF}}=3$. We see that that the FRG result is a considerable improvement over the scaling function in MF approximation, as it is much closer to the MC result, giving us good confidence in the quality of our truncation \eqref{eq:freeEnergyTrunc}.

\begin{figure*}[t]
    \centering
    \begin{minipage}{0.34\linewidth}
        \centering
        {(a)}
        \includegraphics[width=\linewidth]{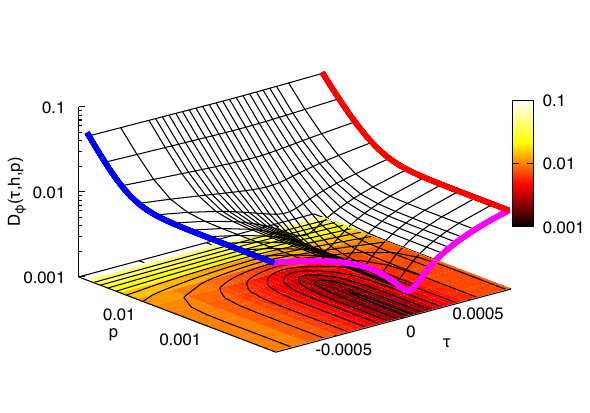}
    \end{minipage}
    \begin{minipage}{0.32\linewidth}
        \centering
        {(b)}
        \includegraphics[width=\linewidth]{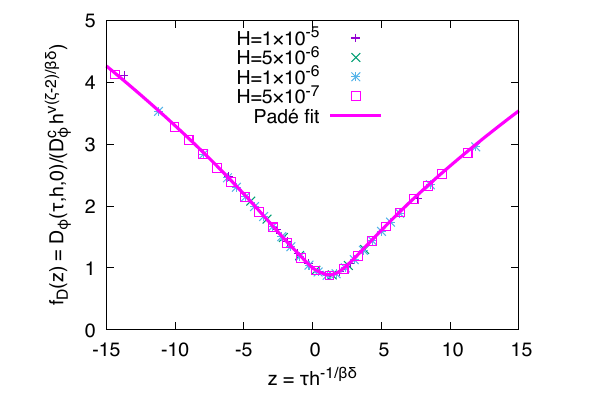}
    \end{minipage}
    \begin{minipage}{0.32\linewidth}
        \centering
        {(c)}
        \includegraphics[width=\linewidth]{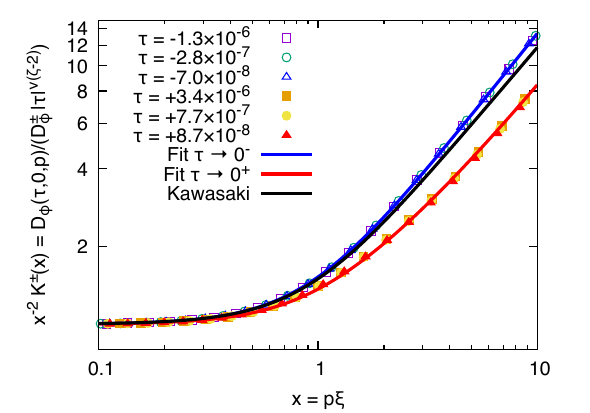}
    \end{minipage}
    \caption{(a) Thermal diffusivity $D_{\phi}(\tau,h,p)$ at fixed $h \approx 1.67\times 10^{-7}$ 
    as a function of reduced temperature $\tau$ and wavenumber $p$ within the scaling region. (b) Universal function $f_{D}(z)$ for the magnetic scaling of the thermal diffusivity $D_{\phi}(\tau,h,0)$ at $p = 0$. (c) Scaling functions $K^{\pm}(x)$ for the universal temperature and wavenumber dependence of the thermal diffusivity $D_{\phi}(\tau,0,p)$ obtained when the critical point is approached from the symmetric phase $\tau \to 0^{+}$ (red) or the broken phase $\tau \to 0^{-}$ (blue) at $h=0$. \label{fig:Dp_scaling}}
\end{figure*}
\begin{figure*}[t]
    \centering
    \begin{minipage}{0.34\linewidth}
        \centering
        {(a)}
        \includegraphics[width=\linewidth]{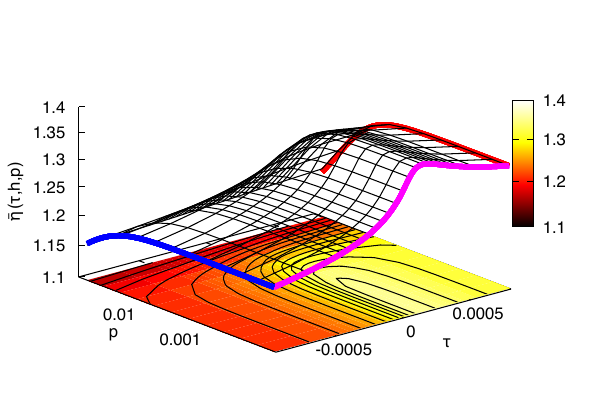}
    \end{minipage}
    \begin{minipage}{0.32\linewidth}
        \centering
        {(b)}
        \includegraphics[width=\linewidth]{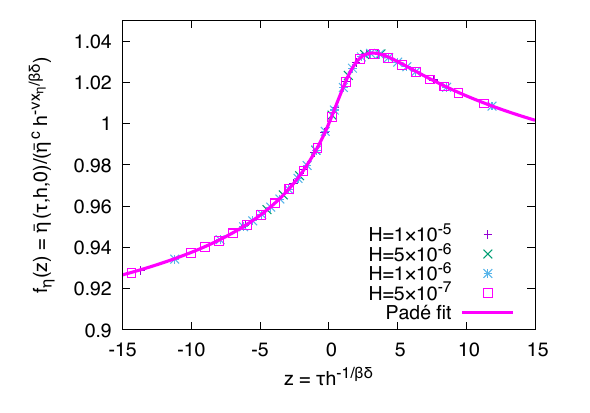}
    \end{minipage}
    \begin{minipage}{0.32\linewidth}
        \centering
        {(c)}
        \includegraphics[width=\linewidth]{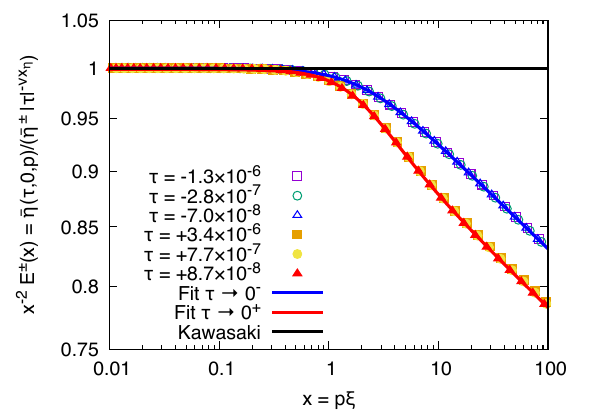}
    \end{minipage}
    \caption{(a) Shear viscosity $\bar{\eta}(\tau,h,p)$ at fixed $h \approx 1.67\times 10^{-7}$ 
    as a function of reduced temperature $\tau$ and wavenumber $p$ within the scaling region. (b) Universal function $f_{\eta}(z)$ for the magnetic scaling of the shear viscosity $\bar{\eta}(\tau,h,0)$ at $p = 0$. (c) Scaling functions $E^{\pm}(x)$ for the universal temperature and wavenumber dependence of the shear viscosity $\bar{\eta}(\tau,0,p)$ obtained when the critical point is approached from the symmetric phase $\tau \to 0^{+}$ (red) or the broken phase $\tau \to 0^{-}$ (blue) at $h=0$. \label{fig:eta_scaling}}
\end{figure*}

\paragraph{Dynamic critical behavior.} Near the critical point the transport coefficients admit universal scaling as well \cite{Hohenberg:1977ym}. These are the thermal diffusivity $D_{\phi}(p) \equiv \gamma_{\phi}(p)/(p^2 \chi_{\phi}(p))$, with the static susceptibility $\chi_{\phi}(p) = (2U'(\phi_0^2)+4\phi_0^2 U''(\phi_0^2)+Z(p)p^2)^{-1}$ at the minimum $\phi_0$ of the full free energy, and the shear viscosity $\bar{\eta}(p) \equiv \gamma_j(p)/p^2$. 
Our FRG results for the transport coefficients $D_{\phi}(\tau,h,p)$ and $\bar{\eta}(\tau,h,p)$ as a function of reduced temperature $\tau$ and wavenumber $p$ at a rather small external field $h \approx 1.67\times 10^{-7}$ 
are shown in Figs.~\ref{fig:Dp_scaling} (a) and \ref{fig:eta_scaling} (a), where the range of parameters has been chosen such that the universal singular part dominates over the regular part. The plots on the corresponding right-hand sides show slices of these two-dimensional scaling functions.

We first focus on the thermal diffusivity $D_{\phi}$,  
which is expected to vanish near the critical point as $D_{\phi} \sim \xi^{2-\zeta}$ with $\zeta \approx 3$ in three dimensions, where $\zeta$ denotes the dynamic critical exponent of Model~H. To see this, note that the equation of motion \eqref{eq:eomsHTrunc} 
entails that the relaxation rate of 
fluctuations with critical wavenumber $p\sim \xi^{-1}$ is $\Gamma_{\xi} \sim D_{\phi}/\xi^{2}$. 
Comparing with the expectation from critical slowing down $\Gamma_{\xi} \sim \xi^{-\zeta}$ directly yields $D_{\phi} \sim \xi^{2-\zeta}$.

Analogous to the Widom-Griffiths scaling in \eqref{eq:phiScaling}, the magnetic scaling of $D_{\phi}$ can be described by a scaling function $f_D(z)$,
\begin{align}
    D_{\phi}(\tau,h,0) &=  D_{\phi}^c h^{\frac{\nu}{\beta\delta}(\zeta-2)}
    f_D(z) \,, \label{eq:DpMagneticScaling}
\end{align}
whose normalization 
is given by $f_D(0) = 1$. Using the static critical exponent $\nu/\beta\delta = (1+1/\delta)/d \approx 0.409$ from \mbox{(hyper-)scaling} relations with  $\beta$ and $\delta$ as specified above,
we obtaind for the dynamic critical exponent $\zeta \approx 3.06$, which is consistent with the result $\zeta = 3.013(58)$ from numerical simulations \cite{Chattopadhyay:2024jlh}. Our FRG result for the scaling function $f_D(z)$ is shown in Fig.~\ref{fig:Dp_scaling} (b), which we fit with a function that is based on Pad\'e approximants and ensures the correct asymptotic behavior for $z \to \pm\infty$ \cite{SuppMat}.

When approaching the critical point at fixed $h=0$, the dynamic scaling hypothesis entails that  the wavenumber $p$ dependent thermal diffusivity scales with reduced temperature $\tau$ as \cite{Hohenberg:1977ym} 
\begin{align}
    D_{\phi}(\tau,0,p) = D_{\phi}^{\pm} |\tau| ^{-\nu(2-\zeta)} \frac{K^{\pm}(p\xi(\tau,0))}{(p\xi(\tau,0))^2} \,.
    \label{eq:DpTauPScaling}
\end{align}
The correlation length $\xi(\tau,h)$ has been introduced here  to render the argument of $K^{\pm}$ dimensionless, with $\xi(\tau,0) = f^{\pm} |\tau|^{-\nu}$ for $\tau \to 0^{\pm}$
\footnote{In practice, we employ the definition of the correlation length that is based on the second moment of the static susceptibility, as discussed  in the Supplemental Material \cite{SuppMat}.}. The non-universal amplitude $D_{\phi}^{\pm}$ is fixed by requiring $K^{\pm}(x)/x^2 \to 1$, for $x \to 0$, in the commonly defined scaling function $K^\pm(x)$ of the relaxation rate, where
the superscript $\pm$ indicates the approach to criticality from the symmetric ($+$) and the broken ($-$) phase, respectively. 
Strikingly, if one neglects the renormalisation of the shear viscosity and assumes the mean-field form for the static susceptibility of the order parameter, one re-derives a closed analytic form for the scaling function (see Supplemental Material~\cite{SuppMat})
\begin{align}
    K^{\pm}(x) = \frac{3}{4} (1+x^2+(x^3-x^{-1})\arctan x) \label{eq:KawFnc}
\end{align}
known as the Kawasaki approximation~\cite{Kawasaki:1970dpc}, where within this approximation, the scaling function $K$ becomes independent of the thermodynamic path $(z)$ towards the critical point. Our FRG results are shown Fig.~\ref{fig:Dp_scaling} (c). The Kawasaki approximation  \eqref{eq:KawFnc} is shown as a black line, which lies between the FRG results for the symmetric phase (red) and the broken phase (blue).

We now turn to the scaling of the shear viscosity $\bar{\eta} \sim \xi^{-x_{\eta}}$, which is neglected in the Kawasaki approximation. The magnetic scaling of the shear viscosity 
can be once more described by a scaling function,
\begin{equation}
    \bar{\eta}(\tau,h,0) = \bar{\eta}^c h^{-\frac{\nu x_{\eta}}{\beta\delta}} f_{\eta}(z) \label{eq:etaMagneticScaling}
\end{equation}
with $f_{\eta}(0) = 1$.
Our rescaled FRG results are shown in Fig.~\ref{fig:eta_scaling} (b). Our result for $x_{\eta} \approx 0.047$ obtained from the magnetic scaling of the shear viscosity $\bar{\eta}(0,h,0) \sim h^{-\frac{\nu x_{\eta}}{\beta\delta}}$ for $h\to 0^+$ appears to deviate from the experimental value $x_{\eta}=0.0690(6)$ measured in the CVX experiment \cite{PhysRevE.60.4079}. However, if we evaluate the flow of the kinetic coefficients at the infrared (IR) minimum instead of the $k$-dependent minimum \cite{SuppMat}, we obtain a larger value $x_{\eta} \approx 0.06$ in the symmetric phase, which is an indication that the experimental value is within the remaining systematic error of the truncated real-time FRG flow.

Finally, the wavenumber dependent shear viscosity scales with reduced temperature as \cite{Hohenberg:1977ym}
\begin{align}
    \bar{\eta}(\tau,0,p) =  \bar{\eta}^{\pm} |\tau|^{-\nu x_{\eta}} \frac{E^{\pm}(p\xi(\tau,0))}{(p\xi(\tau,0))^2}
\end{align}
where the scaling functions are normalized according to $E^{\pm}(x)/x^2 \to 1$ for $x\to 0$. Our real-time FRG results are shown in
Fig.~\ref{fig:eta_scaling} (c), which we 
observe to deviate from the Kawasaki approximation by $\lesssim 20\%$ for the range of $x$ shown.

\emph{Comparison with experiment \& error estimation.} Experimentally, the order-parameter relaxation rate has been measured in a variety of fluids near the liquid-gas critical point \cite{Swinney:1973zz}. Instead of considering the Kawasaki function, one usually considers the scaling function $\Omega^{\pm}(x) = 6\pi R^{\pm} x^{-3} K^{\pm}(x)$ which can be experimentally extracted as
\begin{align}
    D_{\phi}(\tau,0,p) = \frac{g^2 T_c p}{ 6\pi\bar{\eta}(\tau,0,0)} \, \Omega^{\pm}(p\xi(\tau,0))  \label{eq:GamPrime}
\end{align}
by measuring the correlation length $\xi(\tau,0)$, the temperature dependence of the shear-viscosity $\bar{\eta}(\tau,0,0)$ and the temperature and wavenumber dependence of the thermal diffusivity $D_{\phi}(\tau,0,p)$~ \cite{Swinney:1973zz}. In the above definition $R^{\pm}$ denote the universal amplitude ratios \cite{PhysRevB.13.2110}
\begin{equation}
    R^{\pm} \equiv \lim_{\tau\to 0^{\pm}} \frac{\bar{\eta}(\tau,0,0) D_{\phi}(\tau,0,0) \xi(\tau,0)}{g^2 T_c}
\end{equation}
which  in the Kawasaki approximation~\eqref{eq:KawFnc} are identically given by $R^{+} = R^- = 1/6\pi$~\cite{SuppMat}.

\begin{figure}[t]
    \centering
    \includegraphics[width=\linewidth]{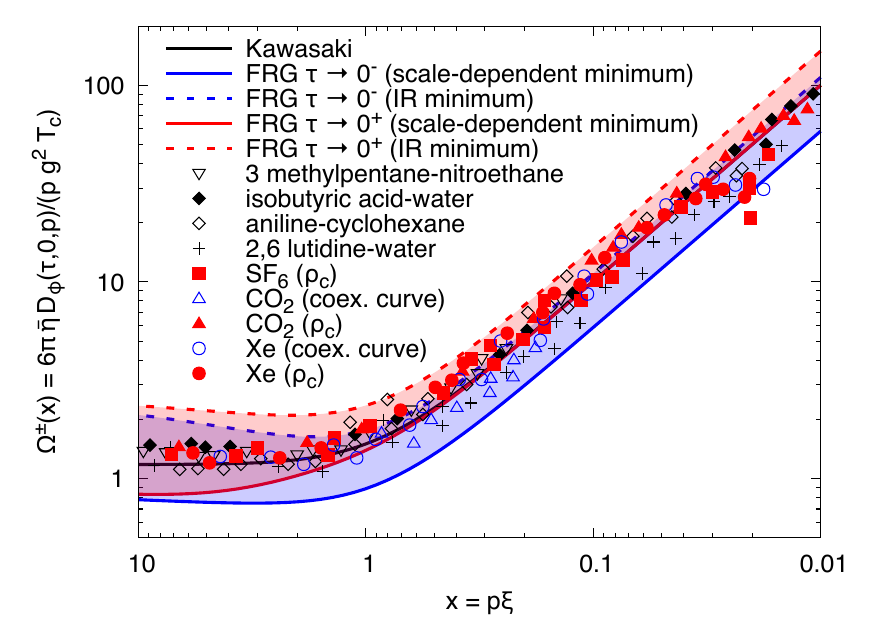}
    \caption{ Comparison of our FRG results for $\Omega^{\pm}(x)$ to experimental data of various fluids \cite{Swinney:1973zz}. The critical isochore of the fluid corresponds to $h=0,\tau\to 0^+$ (indicated as red), and coexistence curve corresponds to $h=0,\tau\to 0^-$ (indicated as blue), respectively.  
    \label{fig:kawasaki_exp_comp}}
\end{figure}

In Fig.~\ref{fig:kawasaki_exp_comp} we present a comparison of our real-time FRG results for $\Omega^{\pm}(x)$, shown as solid lines, to the experimental data compiled in~\cite{Swinney:1973zz}. Approaching the critical point from above (below) at $h=0$ corresponds to the critical isochore (the coexistence curve) in the fluid. The corresponding data points in Fig.~\ref{fig:kawasaki_exp_comp} are indicated as red (blue). One can see that our results describe the overall qualitative shape of the data points quite well, but -- unlike the Kawasaki approximation -- the real-time FRG provides different scaling functions depending on whether the critical point is approached from above or below. 

Finally, to estimate systematic errors of our real-time FRG calculation, we evaluate the flow of the kinetic coefficients at the fixed minimum of the free energy in the IR limit $k\to 0$ \cite{SuppMat}. The corresponding results are shown as dashed lines in Fig.~\ref{fig:kawasaki_exp_comp}. 
Since an exact solution of the FRG flow would not depend on the expansion point in field space, the difference to the expansion around the $k$-dependent minimum (shown as the two bands in Fig.~\ref{fig:kawasaki_exp_comp}) can be regarded as an estimate for the systematic error induced by the truncation of the FRG flow, which we observe to be roughly of the same order of magnitude as the spread of the experimental data in the hydrodynamic regime $x \lesssim 1$. On the other hand, in the critical regime $x \gtrsim 1$ the systematic error of the FRG appears to be larger than the spread of the experimental data, indicating that the present truncation needs further improvement in this regime. 

\paragraph{Conclusion and outlook.} In this work, we have studied the universal scaling of the thermal diffusivity and the shear viscosity near the liquid-gas critical point by solving a truncated real-time FRG flow of Model~H \cite{Roth:2024rbi,Roth:2024hcu}. We have mapped out the universal dependence of these transport coefficients on temperature, external (magnetic) field, and wavenumber, and provided fit functions for various one-dimensional slices. We have compared our FRG results (including an estimate of the systematic error) to experimental data from various fluids \cite{Swinney:1973zz}. In particular, the real-time FRG provides scaling functions which show a (mild) dependence on the thermodynamic path towards the critical point, unlike the one obtained in Kawasaki approximation \cite{Kawasaki:1970dpc}. Such a dependence would be consistent with the experimental data of Ref.~\cite{Swinney:1973zz}, motivating further investigation.

We emphasize that this work lays the foundation for future FRG studies, employing more sophisticated truncation schemes, which have been successfully employed in quantitative computations of static critical exponents \cite{Balog:2019rrg}. 
Clearly, an important first step would be to simply improve the truncation of the static flow to higher-orders in the derivative expansion, whereby the resulting anomalous dimension $\eta = 0.03648$ at order $\mathcal{O}(\partial^6)$ \cite{Balog:2019rrg} is known to be much closer to the high-precision value $\eta=0.0362978(20)$ from conformal bootstrap \cite{Kos:2016ysd}. 

\paragraph{Acknowledgements.}
This work was supported by the Deutsche Forschungsgemeinschaft (DFG, German Research Foundation) through the CRC-TR 211 ‘Strong-interaction matter under extreme conditions’-project number 315477589 – TRR 211.  Computational resources were provided by the HPC Core Facility of Justus-Liebig University Giessen. 

\bibliographystyle{h-physrev3}
\bibliography{refs}

\begin{thebibliography}{10}

\bibitem{10.1063/1.3088050}
M.~L. Huber {\em et~al.},
\newblock J. Phys. Chem. Ref. Data {\bf 38}, 101 (2009).

\bibitem{10.1063/1.4738955}
M.~L. Huber {\em et~al.},
\newblock J. Phys. Chem. Ref. Data {\bf 41}, 033102 (2012).

\bibitem{PhysRevE.60.4079}
R.~F. Berg, M.~R. Moldover, and G.~A. Zimmerli,
\newblock Phys. Rev. E {\bf 60}, 4079 (1999).

\bibitem{Hohenberg:1977ym}
P.~C. Hohenberg and B.~I. Halperin,
\newblock Rev. Mod. Phys. {\bf 49}, 435 (1977).

\bibitem{Onuki_2002}
A.~Onuki,
\newblock {\em Phase Transition Dynamics} (Cambridge University Press, 2002).

\bibitem{Swinney:1973zz}
H.~L. Swinney and D.~L. Henry,
\newblock Phys. Rev. A {\bf 8}, 2586 (1973).

\bibitem{Stephanov:2017ghc}
M.~Stephanov and Y.~Yin,
\newblock Phys. Rev. D {\bf 98}, 036006 (2018), 1712.10305.

\bibitem{Stephanov:1998dy}
M.~A. Stephanov, K.~Rajagopal, and E.~V. Shuryak,
\newblock Phys. Rev. Lett. {\bf 81}, 4816 (1998), hep-ph/9806219.

\bibitem{Stephanov:2011pb}
M.~A. Stephanov,
\newblock Phys. Rev. Lett. {\bf 107}, 052301 (2011), 1104.1627.

\bibitem{Berdnikov:1999ph}
B.~Berdnikov and K.~Rajagopal,
\newblock Phys. Rev. D {\bf 61}, 105017 (2000), hep-ph/9912274.

\bibitem{Son:2004iv}
D.~T. Son and M.~A. Stephanov,
\newblock Phys. Rev. D {\bf 70}, 056001 (2004), hep-ph/0401052.

\bibitem{Kawasaki:1970dpc}
K.~Kawasaki,
\newblock Annals Phys. {\bf 61}, 1 (1970).

\bibitem{PhysRevLett.29.48}
K.~Kawasaki and S.-M. Lo,
\newblock Phys. Rev. Lett. {\bf 29}, 48 (1972).

\bibitem{10.1143/PTP.55.1384}
T.~Ohta and K.~Kawasaki,
\newblock Prog.\ Theor.\ Phys. {\bf 55}, 1384 (1976).

\bibitem{10.1143/PTP.64.536}
T.~Ohta,
\newblock Prog.\ Theor.\ Phys. {\bf 64}, 536 (1980).

\bibitem{Balog:2019rrg}
I.~Balog, H.~Chat{\'e}, B.~Delamotte, M.~Marohnic, and N.~Wschebor,
\newblock Phys. Rev. Lett. {\bf 123}, 240604 (2019), 1907.01829.

\bibitem{Roth:2024rbi}
J.~V. Roth, Y.~Ye, S.~Schlichting, and L.~von Smekal,
\newblock JHEP {\bf 01}, 118 (2025), 2403.04573.

\bibitem{Roth:2024hcu}
J.~V. Roth, Y.~Ye, S.~Schlichting, and L.~von Smekal,
\newblock Phys. Rev. D {\bf 111}, L111901 (2025), 2409.14470.

\bibitem{Chen:2024lzz}
Y.-r. Chen, Y.-y. Tan, and W.-j. Fu,
\newblock Phys. Rev. D {\bf 111}, 094025 (2025), 2406.00679.

\bibitem{Florio:2021jlx}
A.~Florio, E.~Grossi, A.~Soloviev, and D.~Teaney,
\newblock Phys. Rev. D {\bf 105}, 054512 (2022), 2111.03640.

\bibitem{Chattopadhyay:2024jlh}
C.~Chattopadhyay, J.~Ott, T.~Schaefer, and V.~V. Skokov,
\newblock Phys. Rev. Lett. {\bf 133}, 032301 (2024), 2403.10608.

\bibitem{PhysRevB.13.1299}
B.~I. Halperin, P.~C. Hohenberg, and E.~D. Siggia,
\newblock Phys. Rev. B {\bf 13}, 1299 (1976).

\bibitem{Wetterich:1992yh}
C.~Wetterich,
\newblock Phys. Lett. B {\bf 301}, 90 (1993), 1710.05815.

\bibitem{DePolsi:2020pjk}
G.~De~Polsi, I.~Balog, M.~Tissier, and N.~Wschebor,
\newblock Phys. Rev. E {\bf 101}, 042113 (2020), 2001.07525.

\bibitem{Murgana:2023xrq}
F.~Murgana, A.~Koenigstein, and D.~H. Rischke,
\newblock Phys. Rev. D {\bf 108}, 116016 (2023), 2303.16838.

\bibitem{Note1}
Notice that the Langevin noise terms have disappeared since the coarse-grained
  equations of motion hold for the expectation values of the fields (which we
  here denote by the same symbols $\phi $ and $\protect \boldsymbol {j}$ for
  simplicity).

\bibitem{SuppMat}
See Supplemental Material for details.

\bibitem{Note2}
We employ the initial conditions $m^2_{\Lambda }=-1$, $\lambda _{\Lambda }=1$,
  $w=1$ at the UV scale $\Lambda =\pi $. Power-law fits in the scaling regime
  of \protect \eqref {eq:phiAtTc} and \protect \eqref {eq:phiAtH0} yield the
  non-universal amplitudes $B^c/H_0^{1/\delta } \approx 2.03$ and $B \approx
  2.60$. The non-universal amplitude $H_0$ is fixed by the normalization
  $f_G(z)/(-z)^{\beta } \to 1$ for $z \to -\infty $, which is equivalent to
  $B=B^c$ and yields $H_0=(B/(B^c/H_0^{1/\delta }))^{\delta }\approx 2.99$.

\bibitem{Karsch:2023pga}
F.~Karsch, M.~Neumann, and M.~Sarkar,
\newblock Phys. Rev. D {\bf 108}, 014505 (2023), 2304.01710.

\bibitem{Note3}
In practice, we employ the definition of the correlation length that is based
  on the second moment of the static susceptibility, as discussed in the
  Supplemental Material \cite {SuppMat}.

\bibitem{PhysRevB.13.2110}
E.~D. Siggia, B.~I. Halperin, and P.~C. Hohenberg,
\newblock Phys. Rev. B {\bf 13}, 2110 (1976).

\bibitem{Kos:2016ysd}
F.~Kos, D.~Poland, D.~Simmons-Duffin, and A.~Vichi,
\newblock JHEP {\bf 08}, 036 (2016), 1603.04436.

\bibitem{DZYALOSHINSKII198067}
I.~Dzyaloshinskii and G.~Volovick,
\newblock Annals of Physics {\bf 125}, 67 (1980).

\bibitem{Berges:2012ty}
J.~Berges and D.~Mesterhazy,
\newblock Nucl. Phys. B Proc. Suppl. {\bf 228}, 37 (2012), 1204.1489.

\bibitem{Huelsmann:2020xcy}
S.~Huelsmann, S.~Schlichting, and P.~Scior,
\newblock Phys. Rev. D {\bf 102}, 096004 (2020), 2009.04194.

\bibitem{Huber:2019dkb}
M.~Q. Huber, A.~K. Cyrol, and J.~M. Pawlowski,
\newblock Comput. Phys. Commun. {\bf 248}, 107058 (2020), 1908.02760.

\bibitem{PhysRevB.10.2818}
M.~E. Fisher and A.~Aharony,
\newblock Phys. Rev. B {\bf 10}, 2818 (1974).

\bibitem{Pelissetto:2000ek}
A.~Pelissetto and E.~Vicari,
\newblock Phys. Rept. {\bf 368}, 549 (2002), cond-mat/0012164.

\end{thebibliography}

\renewcommand{\thesubsection}{{S.\arabic{subsection}}}
\setcounter{section}{0}

\onecolumngrid

\section*{Supplemental Material}

\subsection{Static FRG flow for the Landau-Ginzburg-Wilson model}

\subsubsection{Flow of effective potential}

With the external field $H$ included, the total effective potential in \eqref{eq:freeEnergyTrunc} has the form (with $\rho \equiv \phi^2$)
\begin{align}
    V_k(\phi) = U_k(\rho) - H\phi \,. \label{eq:effPotWithH}
\end{align}
The flow equation for the $Z_2$-invariant part $U_k(\rho)$ is obtained by evaluating the Wetterich equation \eqref{eq:flowOfFk} for a homogeneous background field expectation value,
\begin{align}
    \partial_k U_k(\rho) = \frac{T}{2} \int \frac{d^dq}{(2\pi)^d} \frac{\partial_k R_k^{\phi}(q)}{2U_k'(\rho) + 4\rho U_k''(\rho) + Z_k(q)q^2 + R_k^{\phi}(q)} \,.
\end{align}
In practice, we expand $U_k(\rho)$ around the minimum of $V_k(\phi)$ up to some finite order $N_{\text{ord}}$, 
\begin{align}
    U_k(\rho) = \sum_{n=1}^{N_{\text{ord}}} \frac{\lambda_{n,k}}{n!} (\rho-\rho_{0,k})^n \,. \label{eq:effPotTaylorExpFiniteH}
\end{align}
In particular, minimizing the effective potential \eqref{eq:effPotWithH} fixes $\lambda_{1,k} = H/(2\sqrt{\rho_{0,k}})$. 
The flow equation for $\rho_{0,k}$ is given by
\begin{align}
    \partial_k \rho_{0,k}=-\frac{\partial_k U_k'(\rho_{0,k})}{\lambda_{2,k}+\frac{H}{4}\rho_{0,k}^{-\frac{3}{2}}} \label{eq:flowOfRho0}
\end{align}
with 
\begin{align}
    \partial_k U_k'(\rho_{0,k}) = - \frac{T}{2} \int \frac{d^dq}{(2\pi)^d} \frac{(6 \lambda_{2,k} + 4\rho_{0,k} \lambda_{3,k})\partial_k R_k^{\phi}(q)}{(m_k^2 + Z_k(q)q^2 + R_k^{\phi}(q))^2}
\end{align}
using $m_k^2 \equiv 2\lambda_{1,k} + 4\rho_{0,k} \lambda_{2,k}$. Since the integrand is spherically symmetric, the integrals over the $d-1$ angular coordinates can be evaluated analytically in hyperspherical coordinates,
\begin{align}
    \partial_k U_k'(\rho_{0,k}) = - \frac{T}{2} \frac{S_{d-1}}{(2\pi)^d} \int_0^{\infty}dq\,q^{d-1}\frac{(6 \lambda_{2,k} + 4\rho_{0,k} \lambda_{3,k})\partial_k R_k^{\phi}(q)}{(m_k^2 + Z_k(q)q^2 + R_k^{\phi}(q))^2}
\end{align}
where $S_{d-1} \equiv 2\pi^{d/2}/\Gamma(d/2)$ denotes the surface area of the $(d-1)$-dimensional unit sphere $S^{d-1}$.

Similarly, the flow of the coefficients $\lambda_{n,k}$ is given by
\begin{align}
    \partial_k \lambda_{n,k} = \frac{T}{2} \frac{S_{d-1}}{(2\pi)^d} \int_0^{\infty} \frac{\partial^n}{\partial \rho^n}  \frac{dq\,q^{d-1}\,\partial_k R_k^{\phi}(q)}{2U_k'(\rho) + 4\rho U_k''(\rho) + Z_k(q)q^2 + R_k^{\phi}(q)} \bigg\rvert_{\rho=\rho_{0,k}} + \lambda_{n+1,k} \partial_k \rho_{0,k} \,. \label{eq:effPotTaylorCoeffFlow}
\end{align}

\subsubsection{Flow of wave function renormalization factor}

The flow of the wavenumber dependent ($p$) wave function renormalization factor $Z(p)$ is obtained from the flow of the two-point function $F_k^{(2)}(\vec{p}) \equiv \delta^2 F_k/\delta \phi(-\vec{p})\delta \phi(\vec{p})$, subtracted at $\vec{p}=0$ and evaluated at the $k$-dependent equilibrium field configuration $\phi=\phi_{0,k}$, which yields
\begin{align}
    \partial_k Z_k(p) p^2 &= \partial_k (F_k^{(2)}(p) - F_k^{(2)}(p=0) ) \nonumber \\
    &= \frac{\kappa_k^2 T}{2} \int \frac{d^dq}{(2\pi)^d} \bigg[ \frac{\partial_k R_k^{\phi}(q)}{(m_k^2 + Z_k(q)q^2 + R_k^{\phi}(q))^2 (m_k^2 + Z_k(r)r^2 + R_k^{\phi}(r))} + \nonumber \\
    &\quad \frac{\partial_k R_k^{\phi}(q)}{(m_k^2 + Z_k(q)q^2 + R_k^{\phi}(q))^2 (m_k^2 + Z_k(s)s^2 + R_k^{\phi}(s))} \bigg] - (\cdots)\bigg\rvert_{p=0} \label{eq:flowOfZ}
\end{align}
with $\vec{r} \equiv \vec{p}-\vec{q}$, and $\vec{s} \equiv \vec{p}+\vec{q}$, and the 3-point coupling constant $\kappa_k \equiv 12\rho_{0,k}^{1/2} \lambda_{2,k} + 8 \rho_{0,k}^{3/2} \lambda_{3,k}$. The integral over $\vec{q}$ can be evaluated in hyperspherical coordinates,
\begin{align}
    \int \frac{d^dq}{(2\pi)^d} = \frac{S_{d-2}}{(2\pi)^d} \int_0^\infty dq\,q^{d-1} \int_{-1}^1 d\cos\theta (1-\cos^2\theta)^{(d-3)/2} 
\end{align}
with $\theta$ denoting the angle between $\vec{p}$ and $\vec{q}$. 

\subsubsection{Numerical implementation and UV initial conditions}

To solve the system of differential equations \eqref{eq:flowOfRho0}, \eqref{eq:effPotTaylorCoeffFlow} and \eqref{eq:flowOfZ} numerically, we use a simple forward Euler scheme with step size $dt = -0.0157$ in `RG time' $t = \log(k/\Lambda)$. We perform the integral over $q$ numerically using a Gauss-Legendre quadrature in the dimensionless variable $\log(q/k)$, and the integrals over $\theta$ using a Gauss-Jacobi quadrature in $\cos\theta$. In practice, we evaluate the Taylor expansion of the effective potential \eqref{eq:effPotTaylorExpFiniteH} up to the order $N_{\text{ord}} = 5$.

We use the exponential regulator for the order parameter, 
\begin{align}
    R_k^{\phi}(p) = \frac{p^2}{e^{p^2/k^2}-1} \,,
\end{align}
and set the regulator for the momentum density to zero, $R_k^{j}(p) = 0$, the reason being that the static fluctuations of the momentum density are non-critical anyway, $\langle \vec{j}^2\rangle = wT/V$, and thus do not need to be regulated.

We choose the following initial conditions at the (arbitrarily chosen) UV scale $\Lambda=\pi$, 
\begin{align}
    Z_{\Lambda}(p) = 1\,,\quad U_{\Lambda}(\phi^2) = \frac{m^2}{2}\phi^2 + \frac{\lambda}{4!}\phi^4
\end{align}
with $m^2=-1$, $\lambda=1$, and set the ($k$-independent) mass density to $w=1$. We find the critical point at $T_c \approx 19.04573534$.

\subsection{Real-time FRG flow for Model H}

The basis for the real-time formulation of the FRG is the Martin-Siggia-Rose (MSR) path-integral formulation of the Langevin equations of motion \eqref{eq:eomsH}, in which the (physical) real-time correlation functions of Model~H can be obtained from the generating functional \cite{Roth:2024hcu}
\begin{align}
    Z[H,\tilde{H},\vec{A},\tilde{\vec{A}}] &= \int \mathcal{D}\phi\,\mathcal{D}\tilde{\phi}\,\mathcal{D}\vec{j}\,\mathcal{D}\tilde{\vec{j}}\, \exp\bigg\{ iS[\phi,\tilde{\phi},\vec{j},\tilde{\vec{j}}] + i\int dt\int d^dx \big( \tilde{H} \phi + \tilde{A}_l j_l \big) \; +  \label{eq:genFunc} \\ \nonumber &\hspace{-1.0cm} i\int dt\int d^dx \, \big[ H\big( -\sigma \vec{\nabla}^2 \tilde{\phi} + g\{\phi,j_m\} \mathcal{T}_{mo} \tilde{j}_o \big) + A_l \big(-\bar{\eta} \vec{\nabla}^2 \mathcal{T}_{lo} \tilde{j}_o + g \mathcal{T}_{lm}\{j_m,\phi\} \tilde{\phi}  + g \mathcal{T}_{lm} \{j_m,j_n\} \mathcal{T}_{no} \tilde{j}_o \big) \big] \bigg\}
\end{align}
with MSR action
\begin{align}
    S[\phi,\tilde{\phi},\vec{j},\tilde{\vec{j}}] = \int dt\int d^dx \bigg\{ &-\tilde{\phi}\left(\frac{\partial \phi}{\partial t} - \sigma \vec{\nabla}^2 \frac{\delta F}{\delta \phi} - g\{\phi,j_{l}\} \mathcal{T}_{lm}\frac{\delta F}{\delta j_m} \right) \nonumber \\
    &-\tilde{j}_l \mathcal{T}_{lm} \left( \frac{\partial j_m}{\partial t} -  \bar{\eta} \vec{\nabla}^2 \frac{\delta F}{\delta j_m} - g\{j_m,\phi\} \frac{\delta F}{\delta \phi} - g\{j_m,j_n\} \mathcal{T}_{no} \frac{\delta F}{\delta j_{no}} \right) \nonumber \\
    &- iT \tilde{\phi} \sigma\vec{\nabla}^2 \tilde{\phi} - iT  \tilde{j}_l \bar{\eta} \mathcal{T}_{lm} \vec{\nabla}^2 \tilde{j}_m \bigg\} \, . \label{eq:MSRAction}
\end{align}
Here we have reformulated the equations of motion \eqref{eq:eomsH} using Poisson brackets \cite{DZYALOSHINSKII198067},
\begin{align}
    \{\phi(\vec{x}),j_l(\vec{x}')\} &= \phi(\vec{x}')\frac{\partial}{\partial x'_l} \delta(\vec{x}-\vec{x}') \, ,\\
    \{j_l(\vec{x}),j_m(\vec{x}')\} &= \left[ j_l(\vec{x}') \frac{\partial}{\partial x'_m} \! - \! j_m(\vec{x}) \frac{\partial}{\partial x_l} \right] \! \delta(\vec{x}-\vec{x}') \, ,\nonumber
\end{align}
and we have employed a shorthand notation for convolutions, e.g.,
\begin{align*}
    \mathcal{T}_{lm}\{j_m,\phi\} \frac{\delta F}{\delta \phi} &\equiv \int d^dx' \int d^dx'' \, \mathcal{T}_{lm}(\vec{x},\vec{x}') \, \{j_m(\vec{x}'),\phi(\vec{x}'')\} \, \frac{\delta F}{\delta \phi(\vec{x}'')}
\end{align*}
with the transversal projector in coordinate space,
\begin{equation*}
    \mathcal{T}_{lm}(\vec{x},\vec{x}') = \int \frac{d^dp}{(2\pi)^d} \,e^{i\vec{p}\cdot(\vec{x}-\vec{x}')} \, \Big( \delta_{lm}-\frac{p_l p_m}{\vec{p}^2}\Big) \,.
\end{equation*}
This reformulation has the advantage that the various symmetries of the MSR action discussed in  \cite{Roth:2024rbi} become manifest.  Importantly, the physical sources $H$, $\vec{A}$ introduced on the level of the effective Hamiltonian \eqref{eq:freeEnergy} couple to the `composite' response fields
\begin{align}
    \tilde{\Phi} &\equiv -\sigma\vec{\nabla}^2 \tilde{\phi} + g\{\phi,j_m\} \mathcal{T}_{mo} \tilde{j}_o \, , \label{eq:ModHResponseFieldsBare} \\ \nonumber
    \tilde{J}_l &\equiv   -\bar{\eta}\vec{\nabla}^2 \mathcal{T}_{lo} \tilde{j}_o + g \mathcal{T}_{lm}\{j_m,\phi\} \tilde{\phi} + g \mathcal{T}_{lm} \{j_m,j_n\} \mathcal{T}_{no} \tilde{j}_o \, .
\end{align}

Connected correlation functions can be obtained from
\begin{align}
    W[H,\tilde{H},\vec{A},\tilde{\vec{A}}] = -i\log Z[H,\tilde{H},\vec{A},\tilde{\vec{A}}] \,. \label{eq:defOfW0}
\end{align}
Due to the variety of fields involved, it is useful to introduce the superfield
$\psi \equiv(\phi,j_1,\ldots,j_d)$ and $\tilde{\Psi} \equiv(\tilde{\Phi},\tilde{J}_1,\ldots,\tilde{J}_d)$
and the conjugate supersources
$J \equiv (H,A_1,\ldots,A_d)$ and $\tilde{J}=(\tilde{H},\tilde{A}_1,\ldots,\tilde{A}_d)$.
We denote corresponding superfield indices by Greek letters $\sA,\sB,\ldots=0,1,\ldots,d$. In this notation, \eqref{eq:defOfW0} becomes
\begin{align}
    W[J,\tilde{J}] = -i\log Z[J,\tilde{J}] \label{eq:defOfW}
\end{align}
and the connected one and two-point correlation functions are (with $\underline{x} \equiv (t,\vec{x})$ and  $\underline{x}' \equiv (t',\vec{x}')$)
\begin{align}
    \frac{\delta W[J,\tilde{J}]}{\delta \tilde{J}_{\sA}(\underline{x})} = \psi_{\sA}(\underline{x}) \,, \quad
    \frac{\delta W[J,\tilde{J}]}{\delta J_{\sA}(\underline{x})} = \tilde{\Psi}_{\sA}(\underline{x}) \,, \label{eq:fieldExpValFromW}
\end{align}
and
\begin{align}
    \frac{\delta^2 W[J,\tilde{J}]}{\delta \tilde{J}_{\sA}(\underline{x})\delta \tilde{J}_{\sB}(\underline{x}')} = iF_{\sA\sB,k}(\underline{x},\underline{x}') \,, \qquad
    \frac{\delta^2 W[J,\tilde{J}]}{\delta \tilde{J}_{\sA}(\underline{x})\delta J_{\sB}(\underline{x}')} = G_{\sA\sB,k}^R(\underline{x},\underline{x}') \,, \label{eq:propagatorsFromW} \\ \nonumber
    \frac{\delta^2 W[J,\tilde{J}]}{\delta J_{\sA}(\underline{x})\delta \tilde{J}_{\sB}(\underline{x}')} = G_{\sA\sB,k}^A(\underline{x},\underline{x}') \,, \qquad
    \frac{\delta^2 W[J,\tilde{J}]}{\delta J_{\sA}(\underline{x})\delta J_{\sB}(\underline{x}')} = i\widetilde{F}_{\sA\sB,k}(\underline{x},\underline{x}') \,.
\end{align}
Adding the regulator \eqref{eq:regInFreeEn} to the effective Hamiltonian replaces the MSR action in the generating functional \eqref{eq:genFunc} by
\begin{align}
    S \to S +\Delta S_k \quad\text{with}\quad \Delta S_k = -\int \! dt \int\! \frac{d^dp}{(2\pi)^d} \, \tilde{\Phi}(t,-\vec{p}) R_k^{\phi}(p) \phi(t,\vec{p})
\end{align}
which thus becomes dependent on the FRG scale $k$. Importantly, one can see here that also the regulator (and not only the physical source $H$) couples to the composite response field $\tilde{\Phi}$ in \eqref{eq:ModHResponseFieldsBare}.

The central object in Wetterich's formulation of the FRG is the effective (average) action $\Gamma_k$, which is defined as the Legendre transform of \eqref{eq:defOfW} minus the regulator term $\Delta S_k$,
\begin{align}
    \Gamma_k[\psi,\tilde{\Psi}] \equiv W_k[J,\tilde{J}] - \Delta S_k[\psi,\tilde{\Psi}] - \int d^{d+1}\underline{x} \, \big( \tilde{J}_{\sA}(\underline{x})\psi_{\sA}(\underline{x}) + J_{\sA}(\underline{x}) \tilde{\Psi}_{\sA}(\underline{x}) \big)
\end{align}
where the sources $J,\tilde{J}$
are given implicitly as a function of the fixed field configurations $\psi,\tilde{\Psi}$ through the inversion of \eqref{eq:fieldExpValFromW}. Its flow equation is given by \cite{Berges:2012ty,Roth:2024rbi}
\begin{equation}
    \partial_k \Gamma_k = \frac{i}{2} \Tr[ \partial_k R_k^{\phi} \left( G_{\phi\phi,k}^R+G_{\phi\phi,k}^A \right) ] \label{eq:realTimeFlowEq} 
\end{equation}
where we already have set the regulator for the momentum density to zero, $R_k^j(p) = 0$, and thus only the order-parameter loop appears at tree level. In particular, the propagators $G_{\phi\phi,k}^R$, $G_{\phi\phi,k}^A$ are related to functional derivatives of the effective average action. Denoting second functional derivatives as 
\begin{align}
    \Gamma_{\sA\sB,k}^{\psi\psi}(\underline{x},\underline{x}') &= \frac{\delta^2 \Gamma_k[\psi,\tilde{\Psi}]}{\delta \psi_{\sA}(\underline{x})\delta \psi_{\sB}(\underline{x}')} \,, \qquad
    \Gamma_{\sA\sB,k}^{\tilde{\Psi}\psi}(\underline{x},\underline{x}') = \frac{\delta^2 \Gamma_k[\psi,\tilde{\Psi}]}{\delta \tilde{\Psi}_{\sA}(\underline{x})\delta \psi_{\sB}(\underline{x}')} \,, \\
    \Gamma_{\sA\sB,k}^{\psi\tilde{\Psi}}(\underline{x},\underline{x}') &= \frac{\delta^2 \Gamma_k[\psi,\tilde{\Psi}]}{\delta \psi_{\sA}(\underline{x})\delta \tilde{\Psi}_{\sB}(\underline{x}')} \,, \qquad
    \Gamma_{\sA\sB,k}^{\tilde{\Psi}\tilde{\Psi}}(\underline{x},\underline{x}') = \frac{\delta^2 \Gamma_k[\psi,\tilde{\Psi}]}{\delta \tilde{\Psi}_{\sA}(\underline{x})\delta \tilde{\Psi}_{\sB}(\underline{x}')} \,, \quad
\end{align}
the various propagators in \eqref{eq:propagatorsFromW} are given by \cite{Berges:2012ty}
\begin{subequations}
\begin{align}
    G_k^R &= -\left\{ (\Gamma_k^{\tilde{\Psi}\psi}-R_k) - \Gamma_k^{\tilde{\Psi}\tilde{\Psi}}   (\Gamma_k^{\psi\tilde{\Psi}}-R_k)^{-1}  \Gamma_k^{\psi\psi} \right\}^{-1} \,, \\
    G_k^A &= -\left\{ (\Gamma_k^{\psi\tilde{\Psi}}-R_k) - \Gamma_k^{\psi\psi}   (\Gamma_k^{\tilde{\Psi}\psi}-R_k)^{-1}  \Gamma_k^{\tilde{\Psi}\tilde{\Psi}} \right\}^{-1} \,, \\
    iF_k &= -\left\{ \Gamma_k^{\psi\psi} - (\Gamma_k^{\psi\tilde{\Psi}}-R_k)   (\Gamma_k^{\tilde{\Psi}\tilde{\Psi}})^{-1}  (\Gamma_k^{\tilde{\Psi}\psi}-R_k) \right\}^{-1} \,, \\
    i\widetilde{F}_k &= -\left\{ \Gamma_k^{\tilde{\Psi}\tilde{\Psi}} - (\Gamma_k^{\tilde{\Psi}\psi}-R_k)   (\Gamma_k^{\psi\psi})^{-1}  (\Gamma_k^{\psi\tilde{\Psi}}-R_k) \right\}^{-1} \,. 
\end{align} \label{eq:propsAsInvsOfGam2}%
\end{subequations}
At the (causal) minimum of the effective average action one has $\tilde{\Psi}=0$ and thus $\Gamma_k^{\psi\psi}=0$ vanishes, which means that the propagators simplify to 
\begin{align}
    G_k^R &= -(\Gamma_k^{\tilde{\Psi}\psi}-R_k)^{-1} \,, \qquad iF_k = G_k^R \Gamma_k^{\tilde{\Psi}\tilde{\Psi}} G_k^A \,,\nonumber \\ 
    G_k^A &= -(\Gamma_k^{\psi\tilde{\Psi}}-R_k)^{-1} \,, \qquad 
    i\widetilde{F}_k = 0 \,,  \label{eq:propsAsInvsOfGam2AtMin} 
\end{align} 
Moreover, the fluctuation-dissipation relation in thermal equilibrium implies the following relation for the Fourier transformed propagators, 
\begin{align}
    iF_k(\omega,\vec{p}) = \frac{T}{\omega}(G_k^R(\omega,\vec{p})-G_k^A(\omega,\vec{p})) \,.
\end{align}

\subsubsection{Diagrammatics}
\label{sec:diagrams}
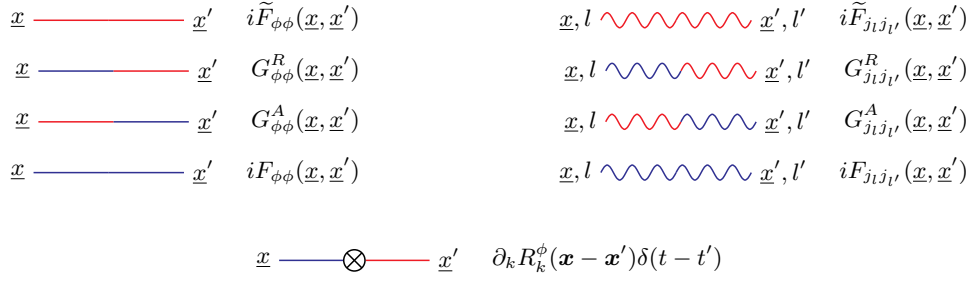
\begin{figure*}
    \centering
    \begin{align*}
    	\begin{split}
			\begin{tikzpicture}[baseline=-0.5ex]
			\draw[pred] (1,0) -- (0,0) node[anchor=east,black] {$\underline{x}$};
		 	\draw[pred] (2,0) node[anchor=west,black] {$\underline{x}'$} -- (1,0);
		\end{tikzpicture} \quad i\widetilde{F}_{\phi \phi}(\underline{x},\underline{x}') \\
		\begin{tikzpicture}[baseline=-0.5ex,samples=120]
		      \draw[pblue] (1,0) -- (0,0) node[anchor=east,black] {$\underline{x}$};
		 	\draw[pred] (2,0) node[anchor=west,black] {$\underline{x}'$} -- (1,0);
		\end{tikzpicture} \quad G^{R}_{\phi \phi}(\underline{x},\underline{x}')
		\\
			\begin{tikzpicture}[baseline=-0.5ex]
			\draw[pred] (1,0) -- (0,0) node[anchor=east,black] {$\underline{x}$};
		 	\draw[pblue] (2,0) node[anchor=west,black] {$\underline{x}'$} -- (1,0);
		\end{tikzpicture} \quad G^{A}_{\phi \phi}(\underline{x},\underline{x}')
		\\
			\begin{tikzpicture}[baseline=-0.5ex]
			\draw[pblue] (1,0) -- (0,0) node[anchor=east,black] {$\underline{x}$};
		 	\draw[pblue] (2,0) node[anchor=west,black] {$\underline{x}'$} -- (1,0);
		\end{tikzpicture} \quad iF_{\phi \phi}(\underline{x},\underline{x}')
	\end{split}
	\begin{split}
		\begin{tikzpicture}[baseline=-0.5ex]
            \draw[pred,smooth,domain=0:1,variable=\x] node[anchor=east,black] {$\underline{x},l$}  plot (\x,{0.1*sin(6*0.5*2*3.14159*\x r)});
            \draw[pred,smooth,domain=0:1,variable=\x] plot (1+\x,{0.1*sin(6*0.5*2*3.14159*(1+\x) r)}) node[anchor=west,black] {$\underline{x}',l'$} ;
		\end{tikzpicture} \quad i\widetilde{F}_{j_l j_{l'}}(\underline{x},\underline{x}') \\
		\begin{tikzpicture}[baseline=-0.5ex,samples=120]
            \draw[pblue,smooth,domain=0:1,variable=\x] node[anchor=east,black] {$\underline{x},l$}  plot (\x,{0.1*sin(6*0.5*2*3.14159*\x r)});
            \draw[pred,smooth,domain=0:1,variable=\x] plot (1+\x,{0.1*sin(6*0.5*2*3.14159*(1+\x) r)}) node[anchor=west,black] {$\underline{x}',l'$} ;
		\end{tikzpicture} \quad G^{R}_{j_l j_{l'}}(\underline{x},\underline{x}')
		\\
		\begin{tikzpicture}[baseline=-0.5ex]
            \draw[pred,smooth,domain=0:1,variable=\x] node[anchor=east,black] {$\underline{x},l$}  plot (\x,{0.1*sin(6*0.5*2*3.14159*\x r)});
            \draw[pblue,smooth,domain=0:1,variable=\x] plot (1+\x,{0.1*sin(6*0.5*2*3.14159*(1+\x) r)}) node[anchor=west,black] {$\underline{x}',l'$} ;
		\end{tikzpicture} \quad G^{A}_{j_l j_{l'}}(\underline{x},\underline{x}')
		\\
		\begin{tikzpicture}[baseline=-0.5ex]
            \draw[pblue,smooth,domain=0:1,variable=\x] node[anchor=east,black] {$\underline{x},l$}  plot (\x,{0.1*sin(6*0.5*2*3.14159*\x r)});
            \draw[pblue,smooth,domain=0:1,variable=\x] plot (1+\x,{0.1*sin(6*0.5*2*3.14159*(1+\x) r)}) node[anchor=west,black] {$\underline{x}',l'$} ;
		\end{tikzpicture} \quad iF_{j_l j_{l'}}(\underline{x},\underline{x}')
	\end{split}
    \end{align*}
    \begin{align*}
    \begin{split}
        \begin{tikzpicture}[baseline=-0.5ex,samples=120]
			\draw[pblue] (1,0) -- (0,0) node[anchor=east,black] {$\underline{x}$};
		 	\draw[pred] (2,0) node[anchor=west,black] {$\underline{x}'$} -- (1,0);
            \filldraw[white] (1,0) circle (0.1414213562);
	        \draw (1,0) circle (0.1414213562);
		 	\draw[black] (1-0.1,-0.1) -- (1+0.1,+0.1);
		 	\draw[black] (1-0.1,+0.1) -- (1+0.1,-0.1);
		\end{tikzpicture} \quad \partial_k R^{\phi}_{k} (\vec{x}-\vec{x}') \delta(t-t')
    \end{split}
    \end{align*}
    \caption{Diagrammatic representations of the various propagators and the regulator derivative. Color indicates the type of the fields in the MSR formalism, i.e.,~blue denotes the classical fields $\phi$, $\vec{j}$, and red the corresponding (composite) response fields $\tilde{\Phi}$, $\tilde{\vec{J}}$. A thick line denotes the sum over superfield indices, i.e.,~over the types of field propagators $\phi$ and $\vec{j}$. In addition, a green line denotes the sum over red and blue, i.e.,~in total over all color permutations.}
    \label{fig:diagrammatics}
\end{figure*}
With the notation of Fig.~\ref{fig:diagrammatics}, the flow equation \eqref{eq:realTimeFlowEq} can be compactly expressed as 
\begin{equation}
    \partial_k \Gamma_k
    = \frac{i}{2} \bigg\{
    \;\begin{tikzpicture}[baseline=-0.5ex]
        \draw[pblue] plot[domain=-90:90, samples=200, smooth, variable=\a] ({0.5*cos(\a)},{0.5*sin(\a)});
        \draw[pred] plot[domain=90:270, samples=200, smooth, variable=\a] ({0.5*cos(\a)},{0.5*sin(\a)});
        \filldraw[white] (0,0.5) circle (0.1414213562);
        \draw (0,0.5) circle (0.1414213562);
        \draw[black] (-0.1,0.5-0.1) -- (+0.1,0.5+0.1);
        \draw[black] (-0.1,0.5+0.1) -- (+0.1,0.5-0.1);
    \end{tikzpicture}\;  +
    \;\begin{tikzpicture}[baseline=-0.5ex]
        \draw[pred] plot[domain=-90:90, samples=200, smooth, variable=\a] ({0.5*cos(\a)},{0.5*sin(\a)});
        \draw[pblue] plot[domain=90:270, samples=200, smooth, variable=\a] ({0.5*cos(\a)},{0.5*sin(\a)});
        \filldraw[white] (0,0.5) circle (0.1414213562);
        \draw (0,0.5) circle (0.1414213562);
        \draw[black] (-0.1,0.5-0.1) -- (+0.1,0.5+0.1);
        \draw[black] (-0.1,0.5+0.1) -- (+0.1,0.5-0.1);
    \end{tikzpicture}\;  \bigg\}
    = \frac{i}{2}
    \;\begin{tikzpicture}[baseline=-0.5ex]
        \draw[pgreen] plot[domain=0:360, samples=200, smooth, variable=\a] ({0.5*cos(\a)},{0.5*sin(\a)});
        \filldraw[white] (0,0.5) circle (0.1414213562);
        \draw (0,0.5) circle (0.1414213562);
        \draw[black] (-0.1,0.5-0.1) -- (+0.1,0.5+0.1);
        \draw[black] (-0.1,0.5+0.1) -- (+0.1,0.5-0.1);
    \end{tikzpicture}\; \label{eq:flowEqDiagram}
\end{equation}
Functional derivatives of the various propagators \eqref{eq:propagatorsFromW} can be efficiently evaluated diagrammatically \cite{Huelsmann:2020xcy}. For example, the rather lengthy expression
\begin{align}
    \frac{\delta}{\delta \tilde{\Phi}(\underline{z})} G^{R}_{j_l j_m}(\underline{x},\underline{y}) = \int_{\underline{v}\underline{w}} \!\!\Big[ &G^{R}_{j_l \psi_{\sA}}(\underline{x},\underline{v}) \Gamma^{\tilde{\Psi}_{\sA} \tilde{\Phi} \psi_{\sB}}(\underline{v},\underline{z},\underline{w}) G^{R}_{\psi_{\sB} j_m}(\underline{w},\underline{y}) + G^{R}_{j_l \psi_{\sA}}(\underline{x},\underline{v}) \Gamma^{\tilde{\Psi}_{\sA} \tilde{\Phi} \tilde{\Psi}_{\sB}}(\underline{v},\underline{z},\underline{w}) i\widetilde{F}_{\psi_{\sB} j_m}(\underline{w},\underline{y}) \;+ \nonumber \\ 
    & iF_{j_l \psi_{\sA}}(\underline{x},\underline{v}) \Gamma^{\psi_{\sA} \tilde{\Phi} \tilde{\Psi}_{\sB}}(\underline{v},\underline{z},\underline{w}) i\widetilde{F}_{\psi_{\sB} j_m}(\underline{w},\underline{y})  + iF_{j_l \psi_{\sA}}(\underline{x},\underline{v}) \Gamma^{\psi_{\sA} \tilde{\Phi} \psi_{\sB}}(\underline{v},\underline{z},\underline{w}) G^R_{\psi_{\sB} j_m}(\underline{w},\underline{y}) \Big]
\end{align}
would be compactly denoted as
\begin{align*}
    \frac{\delta}{\delta \tilde{\Phi}(\underline{z})} G^{R}_{j_l j_m}(\underline{x},\underline{y}) =
    \begin{tikzpicture}[baseline=-0.5ex,samples=120]
        \draw[pblue,smooth,domain=0:0.8,variable=\x] node[anchor=east,black] {$\underline{x},l$}  plot (\x,{0.1*sin(2*2*3.14159*\x/0.8 r)});
        \draw[pred,smooth,domain=0:0.8,variable=\x] plot (2+\x,{0.1*sin(2*2*3.14159*(1+\x/0.8) r)}) node[anchor=west,black] {$\underline{y},m$} ;
        \draw[pgreen,line width=2mm] (0.8,0) -- (2,0);
        \draw[pred] (1.4,-0.1) -- (1.4,-0.7) node[anchor=west,black] {$\underline{z}$};
    \end{tikzpicture} 
\end{align*}
where the intermediate thick green line denotes the sum over all possible field ($\phi$, $\vec{j}$) and color (red, blue) permutations. This shorthand notation allows us to evaluate functional derivatives of the flow equation \eqref{eq:flowEqDiagram} with a drastically reduced number of diagrams.
For instance, the flow equation for the statistical two-point functions $\Gamma_k^{\tilde{\Phi}\tilde{\Phi}}$ and $\Gamma_k^{\tilde{J}\tilde{J}}$ can be compactly expressed as
\begin{align}
    \partial_k \Gamma_k^{\tilde{\Phi}\tilde{\Phi}}(\underline{x},\underline{x}') &=
    \frac{i}{2}\;\bigg\{ \;
    \begin{tikzpicture}[baseline=-0.5ex]
        \draw[pgreen] plot[domain=0:180, samples=200, smooth, variable=\a] ({0.5*cos(\a)},{0.5*sin(\a)});
        \draw[pgreen, line width=1mm] plot[domain=180:360, samples=200, smooth, variable=\a] ({0.5*cos(\a)},{0.5*sin(\a)});
        \draw[pred] (0,-0.55) -- (-0.5,-0.55) ;
        \draw[pred] (0,-0.55) -- (+0.5,-0.55);
        \node[] at (-0.4,-0.85) {\footnotesize $\underline{x}$};
        \node[] at (+0.4,-0.8) {\footnotesize $\underline{x}'$};
        \filldraw[white] (0,0.5) circle (0.1414213562);
        \draw (0,0.5) circle (0.1414213562);
        \draw[black] (-0.1,0.5-0.1) -- (+0.1,0.5+0.1);
        \draw[black] (-0.1,0.5+0.1) -- (+0.1,0.5-0.1);
    \end{tikzpicture} \;\; + \;\;
    \begin{tikzpicture}[baseline=-0.5ex]
        \draw[pgreen,line width=1mm] plot[domain=0:45, samples=200, smooth, variable=\a] ({0.5*cos(\a)},{0.5*sin(\a)});
        \draw[pgreen] plot[domain=45:135, samples=200, smooth, variable=\a] ({0.5*cos(\a)},{0.5*sin(\a)});
        \draw[pgreen,line width=1mm] plot[domain=135:180, samples=200, smooth, variable=\a] ({0.5*cos(\a)},{0.5*sin(\a)});
        \draw[pgreen,line width=1mm] plot[domain=180:360, samples=200, smooth, variable=\a] ({0.5*cos(\a)},{0.5*sin(\a)});
	 	\draw[pred] (-0.55,0) -- (-0.55-0.5,0);
	 	\draw[pred] (0.55,0) -- (0.55+0.5,0);
        \node[] at (-0.9,-0.3) {\footnotesize $\underline{x}$};
        \node[] at (+0.9,-0.25) {\footnotesize $\underline{x}'$};
        \filldraw[white] (0,0.5) circle (0.1414213562);
        \draw (0,0.5) circle (0.1414213562);
        \draw[black] (-0.1,0.5-0.1) -- (+0.1,0.5+0.1);
        \draw[black] (-0.1,0.5+0.1) -- (+0.1,0.5-0.1);
    \end{tikzpicture} \;\; + \; \;
    \begin{tikzpicture}[baseline=-0.5ex]
        \draw[pgreen,line width=1mm] plot[domain=0:45, samples=200, smooth, variable=\a] ({0.5*cos(\a)},{0.5*sin(\a)});
        \draw[pgreen] plot[domain=45:135, samples=200, smooth, variable=\a] ({0.5*cos(\a)},{0.5*sin(\a)});
        \draw[pgreen,line width=1mm] plot[domain=135:180, samples=200, smooth, variable=\a] ({0.5*cos(\a)},{0.5*sin(\a)});
        \draw[pgreen,line width=1mm] plot[domain=180:360, samples=200, smooth, variable=\a] ({0.5*cos(\a)},{0.5*sin(\a)});
	 	\draw[pred] (-0.55,0) -- (-0.55-0.5,0);
	 	\draw[pred] (0.55,0) -- (0.55+0.5,0);
        \node[] at (-0.9,-0.25) {\footnotesize $\underline{x}'$};
        \node[] at (+0.9,-0.3) {\footnotesize $\underline{x}$};
        \filldraw[white] (0,0.5) circle (0.1414213562);
        \draw (0,0.5) circle (0.1414213562);
        \draw[black] (-0.1,0.5-0.1) -- (+0.1,0.5+0.1);
        \draw[black] (-0.1,0.5+0.1) -- (+0.1,0.5-0.1);
    \end{tikzpicture} \;
    \bigg\} \,, \\
    \partial_k \Gamma_k^{\tilde{J}_l\tilde{J}_{l'}}(\underline{x},\underline{x}') &=
    \frac{i}{2}\;\bigg\{ 
    \begin{tikzpicture}[baseline=-0.5ex]
        \draw[pgreen] plot[domain=0:180, samples=200, smooth, variable=\a] ({0.5*cos(\a)},{0.5*sin(\a)});
        \draw[pgreen,line width=1mm] plot[domain=180:360, samples=200, smooth, variable=\a] ({0.5*cos(\a)},{0.5*sin(\a)});
        \draw[pred,smooth,domain=0:0.5,variable=\x] plot (-\x,{-0.55+0.07*(cos(8*3.14159*\x r)-1)});
        \draw[pred,smooth,domain=0:0.5,variable=\x] plot (\x,{-0.55+0.07*(cos(8*3.14159*\x r)-1)});
        \node[] at (-0.4,-0.95) {\footnotesize $\underline{x},l$};
        \node[] at (+0.4,-0.95) {\footnotesize $\underline{x}',l'$};
        \filldraw[white] (0,0.5) circle (0.1414213562);
        \draw (0,0.5) circle (0.1414213562);
        \draw[black] (-0.1,0.5-0.1) -- (+0.1,0.5+0.1);
        \draw[black] (-0.1,0.5+0.1) -- (+0.1,0.5-0.1);
    \end{tikzpicture} + 
    \begin{tikzpicture}[baseline=-0.5ex]
        \draw[pgreen,line width=1mm] plot[domain=0:45, samples=200, smooth, variable=\a] ({0.5*cos(\a)},{0.5*sin(\a)});
        \draw[pgreen] plot[domain=45:135, samples=200, smooth, variable=\a] ({0.5*cos(\a)},{0.5*sin(\a)});
        \draw[pgreen,line width=1mm] plot[domain=135:180, samples=200, smooth, variable=\a] ({0.5*cos(\a)},{0.5*sin(\a)});
        \draw[pgreen,line width=1mm] plot[domain=180:360, samples=200, smooth, variable=\a] ({0.5*cos(\a)},{0.5*sin(\a)});
        \draw[pred,smooth,domain=0:0.5,variable=\x] plot (-\x-0.55,{-0.07*sin(8*3.14159*\x r)});
        \draw[pred,smooth,domain=0:0.5,variable=\x] plot (\x+0.55,{-0.07*sin(8*3.14159*\x r)});
        \node[] at (-0.95,-0.35) {\footnotesize $\underline{x},l$};
        \node[] at (+0.95,-0.35) {\footnotesize $\underline{x}',l'$};
        \filldraw[white] (0,0.5) circle (0.1414213562);
        \draw (0,0.5) circle (0.1414213562);
        \draw[black] (-0.1,0.5-0.1) -- (+0.1,0.5+0.1);
        \draw[black] (-0.1,0.5+0.1) -- (+0.1,0.5-0.1);
    \end{tikzpicture}  + 
    \begin{tikzpicture}[baseline=-0.5ex]
        \draw[pgreen,line width=1mm] plot[domain=0:45, samples=200, smooth, variable=\a] ({0.5*cos(\a)},{0.5*sin(\a)});
        \draw[pgreen] plot[domain=45:135, samples=200, smooth, variable=\a] ({0.5*cos(\a)},{0.5*sin(\a)});
        \draw[pgreen,line width=1mm] plot[domain=135:180, samples=200, smooth, variable=\a] ({0.5*cos(\a)},{0.5*sin(\a)});
        \draw[pgreen,line width=1mm] plot[domain=180:360, samples=200, smooth, variable=\a] ({0.5*cos(\a)},{0.5*sin(\a)});
        \draw[pred,smooth,domain=0:0.5,variable=\x] plot (-\x-0.55,{-0.07*sin(8*3.14159*\x r)});
        \draw[pred,smooth,domain=0:0.5,variable=\x] plot (\x+0.55,{-0.07*sin(8*3.14159*\x r)});
        \node[] at (-0.95,-0.35) {\footnotesize $\underline{x}',l'$};
        \node[] at (+0.95,-0.35) {\footnotesize $\underline{x},l$};
        \filldraw[white] (0,0.5) circle (0.1414213562);
        \draw (0,0.5) circle (0.1414213562);
        \draw[black] (-0.1,0.5-0.1) -- (+0.1,0.5+0.1);
        \draw[black] (-0.1,0.5+0.1) -- (+0.1,0.5-0.1);
    \end{tikzpicture} 
    \bigg\} \,.
\end{align}
In the end, we set the fields to their equilibrium values (see Eq.~\eqref{eq:eqFieldConfig} below), where all propagators and vertex functions vanish which are either acausal (such as $i\widetilde{F}$) or not allowed by symmetry.

\subsubsection{Truncation}

The truncation of the effective average (MSR) action is constrained by symmetry of thermal equilibrium and the extended temporal gauge symmetry \cite{Roth:2024rbi,Roth:2024hcu}. Both of these are still satisfied if the kinetic coefficients are promoted to depend on wavevector $\vec{p}$ and the FRG scale $k$, 
\begin{align}
    \Gamma_k = \int dt \int d^d x \bigg\{ &-\tilde{\phi}\left(\frac{\partial \phi}{\partial t} + \gamma_{\phi,k}(-i\vec{\nabla}) \frac{\delta F_k}{\delta \phi} - g\{\phi,j_l\} \mathcal{T}_{lm} \frac{\delta F_k}{\delta j_m} \right) \nonumber \\
    &-\tilde{j}_l \mathcal{T}_{lm} \left( \frac{\partial j_m}{\partial t} + \gamma_{j,k}(-i\vec{\nabla}) \frac{\delta F_k}{\delta j_m} - g\{j_m,\phi\} \frac{\delta F_k}{\delta \phi} - g\{j_m,j_n\} \mathcal{T}_{no} \frac{\delta F_k}{\delta j_o} \right) \nonumber \\
    &+ iT \tilde{\phi} \gamma_{\phi,k}(-i\vec{\nabla}) \tilde{\phi} + iT  \tilde{j}_l \mathcal{T}_{lm} \gamma_{j,k}(-i\vec{\nabla}) \tilde{j}_m \bigg\} \,, \label{eq:effAvgAction}
\end{align}
where $\tilde{\phi}$ and $\tilde{\vec{j}}$ are implicitly given through the inversion of 
\begin{align}
    \tilde{\Phi} &= \gamma_{\phi,k}(-i\vec{\nabla}) \tilde{\phi} + g\{\phi,j_m\} \mathcal{T}_{mo} \tilde{j}_o \, , \label{eq:ModHResponseFields} \\ \nonumber
    \tilde{J}_l &=   \gamma_{j,k}(-i\vec{\nabla}) \mathcal{T}_{lo} \tilde{j}_o + g \mathcal{T}_{lm}\{j_m,\phi\} \tilde{\phi} + g \mathcal{T}_{lm} \{j_m,j_n\} \mathcal{T}_{no} \tilde{j}_o \, .
\end{align} 
In practice, we write \eqref{eq:ModHResponseFields} in Fourier space at fixed time $t$ (suppressing the index $k$),
\begingroup 
\setlength\arraycolsep{10pt} 
\begin{align}
    \begin{pmatrix}
        \tilde{\Phi}(\vec{p})\\
        \tilde{J}_l(\vec{p})
    \end{pmatrix} &=
    \int \frac{d^dq}{(2\pi)^d} 
    \bigg[\underbrace{\begin{pmatrix}
        \gamma_{\phi}(\vec{p})\delta(\vec{p}+\vec{q}) & 0 \\
        0 & \gamma_{j}(\vec{p})\delta(\vec{p}+\vec{q}) \mathcal{T}_{lo}(\vec{q})
    \end{pmatrix}}_{\equiv J_0(\vec{p},\vec{q})} \\ \nonumber
    &-g\underbrace{
    \begin{pmatrix}
        0 &  ip_m \phi(\vec{p}+\vec{q}) \mathcal{T}_{mo}(-\vec{q})   \\
        -  \mathcal{T}_{lm}(\vec{p}) iq_m \phi(\vec{p}+\vec{q}) &  \mathcal{T}_{lm}(\vec{p}) ip_n j_m(\vec{p}+\vec{q}) \mathcal{T}_{no}(-\vec{q}) -  \mathcal{T}_{lm}(\vec{p}) j_n(\vec{p}+\vec{q}) iq_m \mathcal{T}_{no}(-\vec{q})
    \end{pmatrix}}_{\equiv \Delta J(\vec{p},\vec{q})} \bigg]
    \begin{pmatrix}
        \tilde{\phi}(\vec{q})\\
        \tilde{j}_o(\vec{q})
    \end{pmatrix}
\end{align}
\endgroup
and perform the inversion of $J(\vec{p},\vec{q}) \equiv J_0(\vec{p},\vec{q})-g\Delta J(\vec{p},\vec{q})$  using a Neumann series expansion 
\begin{align}
    J^{-1}(\vec{p},\vec{q}) = J_0^{-1}(\vec{p},\vec{q}) &+ g\int \!\frac{d^dr_1}{(2\pi)^d} \frac{d^dr_2}{(2\pi)^d}  \, J_0^{-1}(\vec{p},\vec{r}_1) \Delta J(\vec{r}_1,\vec{r}_2) J_0^{-1}(\vec{r}_2,\vec{q})  \label{eq:NeumannSeries} \\ \nonumber
    &+ g^2\int \!\frac{d^dr_1}{(2\pi)^d}\cdots \frac{d^dr_4}{(2\pi)^d}  \,  J_0^{-1}(\vec{p},\vec{r}_1) \Delta J(\vec{r}_1,\vec{r}_2) J_0^{-1}(\vec{r}_2,\vec{r}_3)\Delta J(\vec{r}_3,\vec{r}_4) J_0^{-1}(\vec{r}_4,\vec{q}) + \cdots
\end{align}
for small inhomogeneous deviations $\Delta\phi(\vec{x})$ and $\Delta\vec{j}(\vec{x})$ of the field expectation values from the $k$-dependent minimum of the effective average action,
\begin{equation}
    \phi=\phi_{0,k}=const \,, \quad \tilde{\Phi}=0\,, \quad \vec{j}=0 \,, \quad \tilde{\vec{J}} = 0 \,. \label{eq:eqFieldConfig}
\end{equation}
From the zeroth order of the Neumann series \eqref{eq:NeumannSeries} we obtain the propagators 
\begin{align*}
    \begin{split}
        i\widetilde{F}_{\phi\phi,k}(\omega,\vec{p}) &= 0 \,, \\
        G^R_{\phi\phi,k}(\omega,\vec{p}) &= -\frac{\gamma_{\phi,k}(p)}{+i\omega - \gamma_{\phi,k}(p)( m_k^2 + Z_k(p) p^2 + R_{k}^{\phi}(p))} \,, \\
        G^A_{\phi\phi,k}(\omega,\vec{p}) &= -\frac{\gamma_{\phi,k}(p)}{-i\omega - \gamma_{\phi,k}(p)( m_k^2 + Z_k(p) p^2 + R_{k}^{\phi}(p))} \,, \\
        iF_{\phi\phi,k}(\omega,\vec{p}) &= \frac{T}{\omega} \left( G^R_{\phi\phi,k}(\omega,\vec{p}) - G^A_{\phi\phi,k}(\omega,\vec{p}) \right) \,,
    \end{split}
    \begin{split}
        i\widetilde{F}_{j_l j_m,k}(\omega,\vec{p}) &= 0 \,, \\
        G^R_{j_l j_m,k}(\omega,\vec{p}) &= -\frac{\gamma_{j,k}(p)\,\mathcal{T}_{lm}(\vec{p})}{+i\omega - \gamma_{j,k}(p)/w}  \,, \\
        G^A_{j_l j_m,k}(\omega,\vec{p}) &= -\frac{\gamma_{j,k}(p)\,\mathcal{T}_{lm}(\vec{p})}{-i\omega - \gamma_{j,k}(p)/w} \,, \\
        iF_{j_l j_m,k}(\omega,\vec{p}) &= \frac{T}{\omega} \left( G^R_{j_l j_m,k}(\omega,\vec{p}) - G^A_{j_l j_m,k}(\omega,\vec{p}) \right) \,.
    \end{split}
\end{align*}
After evaluating the Neumann series \eqref{eq:NeumannSeries} to second order with \textsc{Mathematica}, we perform the rather cumbersome task of computing the relevant 3-point and 4-point vertex functions as functional derivatives of the truncated effective average action \eqref{eq:effAvgAction} using \textsc{DoFun} \cite{Huber:2019dkb}. For the equilibrium field configuration \eqref{eq:eqFieldConfig}, the non-vanishing vertex functions that enter the diagrams in Fig.~\ref{fig:flowOf2PtFnc} below are (with $\underline{p}\equiv(\omega_p,\vec{p})$, $\underline{q}\equiv(\omega_q,\vec{q})$, $\underline{r}\equiv(\omega_r,\vec{r})$ and $\underline{s}\equiv(\omega_s,\vec{s})$)
\begin{align*}
    \Gamma_k^{\tilde{\Phi}\phi j_l}(\underline{p},\underline{q},\underline{r}) &= -g\,\frac{\omega_r \, (\mathcal{T}_{\vec{r}} \vec{p})_l}{\gamma_{j,k}(r) \, \gamma_{\phi,k}(p)} \, , \\
    \Gamma_k^{\tilde{J}_l \phi \phi}(\underline{p},\underline{q},\underline{r}) &= g \, \frac{\omega_q \, (\mathcal{T}_{\vec{p}} \vec{q})_l}{\gamma_{j,k}(p) \, \gamma_{\phi,k}(q)} + g \,\frac{\omega_r \, (\mathcal{T}_{\vec{p}}\vec{r})_l}{\gamma_{j,k}(p) \, \gamma_{\phi,k}(r)} \, ,  \\
    \Gamma_k^{\tilde{\Phi}\tilde{\Phi}\phi\phi}(\underline{p},\underline{q},\underline{r},\underline{s}) &= \frac{2ig^2 T}{\gamma_{\phi,k}(p) \, \gamma_{\phi,k}(q)} \, \left( \frac{\mathcal{T}_{lm}(\vec{p}+\vec{r})}{\gamma_{j,k}(|\vec{p}+\vec{r}|)} + \frac{\mathcal{T}_{lm}(\vec{q}+\vec{r})}{\gamma_{j,k}(|\vec{q}+\vec{r}|)} \right) r_l s_m \, , \\
    \Gamma_k^{\tilde{J}_l \tilde{J}_m \phi \phi}(\underline{p},\underline{q},\underline{r},\underline{s}) &= \frac{2i g^2 T}{\gamma_{j,k}(p) \, \gamma_{j,k}(q)} \left( \, \frac{(\mathcal{T}_{\vec{p}}(\vec{p}+\vec{r}))_l \, (\mathcal{T}_{\vec{q}}(\vec{q}+\vec{s}))_m}{ \gamma_{\phi,k}(|\vec{p}+\vec{r}|)} + \frac{(\mathcal{T}_{\vec{p}}(\vec{p}+\vec{s}))_{l}  \, (\mathcal{T}_{\vec{q}}(\vec{q}+\vec{r}))_{m}}{ \gamma_{\phi,k}(|\vec{q}+\vec{r}|)} \right) \, .
\end{align*}

\subsubsection{Flow equations for kinetic coefficients}

The fluctuation-dissipation theorem entails that the kinetic coefficients $\gamma_{\phi,k}(p)$ and $\gamma_{j,k}(p)$ can be obtained from two functional derivatives of the effective average action with respect to the composite response fields $\tilde{\Phi}$ and $\tilde{J}_l$,
\begin{align}
    \Gamma_k^{\tilde{\Phi}\tilde{\Phi}}(\omega,\vec{p}) = \frac{2iT}{\gamma_{\phi,k}(p)} \,, \qquad
    \Gamma_k^{\tilde{J}_l\tilde{J}_m}(\omega,\vec{p}) = \frac{2iT\,\mathcal{T}_{lm}(\vec{p})}{\gamma_{j,k}(p)} \,,
\end{align}
which means that their flow equations can be written as
\begin{align}
    \partial_k \gamma_{\phi,k}(\vec{p}) = -\frac{\gamma_{\phi,k}^2(\vec{p})}{2iT} \partial_k \Gamma_k^{\tilde{\Phi}\tilde{\Phi}}(0,\vec{p}) \,, \qquad 
    \partial_k \gamma_{j,k}(\vec{p}) =
     -\frac{\gamma_{j,k}^2(\vec{p})}{2iT} \frac{\mathcal{T}_{lm}(\vec{p})}{d-1} \partial_k \Gamma_k^{\tilde{J}_l \tilde{J}_m}(0,\vec{p}) \,,  \label{eq:kinCoeffFlow} 
\end{align}
using $\mathcal{T}_{lm}(\vec{p}) \mathcal{T}_{lm}(\vec{p})=d-1$. Following the graphical rules discussed in Sec.~\ref{sec:diagrams} above, the resulting flow equations for the statistical 2-point functions $\Gamma_k^{\tilde{\Phi}\tilde{\Phi}}$ and $\Gamma_k^{\tilde{J}\tilde{J}}$ are shown in Fig.~\ref{fig:flowOf2PtFnc}.
\begin{figure*}
    \centering
    \begin{align*}
    \partial_k \Gamma_k^{\tilde{\Phi}\tilde{\Phi}}(x,y) &=
    i\;\bigg\{ \frac{1}{2} 
        \begin{tikzpicture}[baseline=-0.5ex]
            \draw[pgreen] plot[domain=0:180, samples=200, smooth, variable=\a] ({0.5*cos(\a)},{0.5*sin(\a)});
            \draw[pblue] plot[domain=180:360, samples=200, smooth, variable=\a] ({0.5*cos(\a)},{0.5*sin(\a)});
            \draw[pred] (0,-0.5) -- (-0.5,-0.5) ;
            \draw[pred] (0,-0.5) -- (+0.5,-0.5);
            \node[] at (-0.4,-0.75) {\footnotesize $x$};
            \node[] at (+0.4,-0.75) {\footnotesize $y$};
            \fill[black] (0,-0.5) circle (0.05);
            \filldraw[white] (0,0.5) circle (0.1414213562);
            \draw (0,0.5) circle (0.1414213562);
            \draw[black] (-0.1,0.5-0.1) -- (+0.1,0.5+0.1);
            \draw[black] (-0.1,0.5+0.1) -- (+0.1,0.5-0.1);
        \end{tikzpicture} + 
        \begin{tikzpicture}[baseline=-0.5ex]
            \draw[pblue] plot[domain=0:45, samples=200, smooth, variable=\a] ({0.5*cos(\a)},{0.5*sin(\a)});
            \draw[pgreen] plot[domain=45:135, samples=200, smooth, variable=\a] ({0.5*cos(\a)},{0.5*sin(\a)});
            \draw[pblue] plot[domain=135:180, samples=200, smooth, variable=\a] ({0.5*cos(\a)},{0.5*sin(\a)});
            \draw[pblue] plot[domain=180:360, samples=200, smooth, variable=\a] ({0.5*cos(\a)*(1.0-0.1*sin(15*\a))},{0.5*sin(\a)*(1.0-0.1*sin(15*\a))});
    	 	\draw[pred] (-0.5,0) -- (-0.5-0.5,0);
    	 	\draw[pred] (0.5,0) -- (0.5+0.5,0);
            \node[] at (-0.9,-0.25) {\footnotesize $x$};
            \node[] at (+0.9,-0.25) {\footnotesize $y$};
		 	\fill[black] (-0.5,0) circle (0.05);
		 	\fill[black] (+0.5,0) circle (0.05);
            \filldraw[white] (0,0.5) circle (0.1414213562);
            \draw (0,0.5) circle (0.1414213562);
            \draw[black] (-0.1,0.5-0.1) -- (+0.1,0.5+0.1);
            \draw[black] (-0.1,0.5+0.1) -- (+0.1,0.5-0.1);
        \end{tikzpicture}  + 
        \begin{tikzpicture}[baseline=-0.5ex]
            \draw[pblue] plot[domain=0:45, samples=200, smooth, variable=\a] ({0.5*cos(\a)},{0.5*sin(\a)});
            \draw[pgreen] plot[domain=45:135, samples=200, smooth, variable=\a] ({0.5*cos(\a)},{0.5*sin(\a)});
            \draw[pblue] plot[domain=135:180, samples=200, smooth, variable=\a] ({0.5*cos(\a)},{0.5*sin(\a)});
            \draw[pblue] plot[domain=180:360, samples=200, smooth, variable=\a] ({0.5*cos(\a)},{0.5*sin(\a)});
    	 	\draw[pred] (-0.5,0) -- (-0.5-0.5,0);
    	 	\draw[pred] (0.5,0) -- (0.5+0.5,0);
            \node[] at (-0.9,-0.25) {\footnotesize $x$};
            \node[] at (+0.9,-0.25) {\footnotesize $y$};
		 	\fill[black] (-0.5,0) circle (0.05);
		 	\fill[black] (+0.5,0) circle (0.05);
            \filldraw[white] (0,0.5) circle (0.1414213562);
            \draw (0,0.5) circle (0.1414213562);
            \draw[black] (-0.1,0.5-0.1) -- (+0.1,0.5+0.1);
            \draw[black] (-0.1,0.5+0.1) -- (+0.1,0.5-0.1);
        \end{tikzpicture} 
    \bigg\} \\
    \partial_k \Gamma_k^{\tilde{J}_l\tilde{J}_m}(x,y) &=
    i\;\bigg\{ \frac{1}{2} 
        \begin{tikzpicture}[baseline=-0.5ex]
            \draw[pgreen] plot[domain=0:180, samples=200, smooth, variable=\a] ({0.5*cos(\a)},{0.5*sin(\a)});
            \draw[pblue] plot[domain=180:360, samples=200, smooth, variable=\a] ({0.5*cos(\a)},{0.5*sin(\a)});
            \draw[pred,smooth,domain=0:0.5,variable=\x] plot (-\x,{-0.5+0.07*(cos(8*3.14159*\x r)-1)});
            \draw[pred,smooth,domain=0:0.5,variable=\x] plot (\x,{-0.5+0.07*(cos(8*3.14159*\x r)-1)});
            \node[] at (-0.4,-0.85) {\footnotesize $x,l$};
            \node[] at (+0.4,-0.9) {\footnotesize $y,m$};
            \fill[black] (0,-0.5) circle (0.05);
            \filldraw[white] (0,0.5) circle (0.1414213562);
            \draw (0,0.5) circle (0.1414213562);
            \draw[black] (-0.1,0.5-0.1) -- (+0.1,0.5+0.1);
            \draw[black] (-0.1,0.5+0.1) -- (+0.1,0.5-0.1);
        \end{tikzpicture} + 
        \begin{tikzpicture}[baseline=-0.5ex]
            \draw[pblue] plot[domain=0:45, samples=200, smooth, variable=\a] ({0.5*cos(\a)},{0.5*sin(\a)});
            \draw[pgreen] plot[domain=45:135, samples=200, smooth, variable=\a] ({0.5*cos(\a)},{0.5*sin(\a)});
            \draw[pblue] plot[domain=135:180, samples=200, smooth, variable=\a] ({0.5*cos(\a)},{0.5*sin(\a)});
            \draw[pblue] plot[domain=180:360, samples=200, smooth, variable=\a] ({0.5*cos(\a)},{0.5*sin(\a)});
            \draw[pred,smooth,domain=0:0.5,variable=\x] plot (-\x-0.5,{-0.07*sin(8*3.14159*\x r)});
            \draw[pred,smooth,domain=0:0.5,variable=\x] plot (\x+0.5,{-0.07*sin(8*3.14159*\x r)});
            \node[] at (-0.9,-0.3) {\footnotesize $x,l$};
            \node[] at (+0.9,-0.35) {\footnotesize $y,m$};
		 	\fill[black] (-0.5,0) circle (0.05);
		 	\fill[black] (+0.5,0) circle (0.05);
            \filldraw[white] (0,0.5) circle (0.1414213562);
            \draw (0,0.5) circle (0.1414213562);
            \draw[black] (-0.1,0.5-0.1) -- (+0.1,0.5+0.1);
            \draw[black] (-0.1,0.5+0.1) -- (+0.1,0.5-0.1);
        \end{tikzpicture} 
        \bigg\}
    \end{align*}
    \caption{Flow of statistical 2-point functions $\Gamma_k^{\tilde{\Phi}\tilde{\Phi}}$ and $\Gamma_k^{\tilde{J}\tilde{J}}$ at the $k$-dependent equilibrium field configuration \eqref{eq:eqFieldConfig}.}
    \label{fig:flowOf2PtFnc}
\end{figure*}
After evaluating the frequency integrals with the residue theorem, we obtain the following expressions for the individual diagrams appearing on the right-hand side of \eqref{eq:kinCoeffFlow} (suppressing the FRG scale $k$),
\begin{alignat*}{3}
    -\frac{i\gamma_{\phi}^2(p)}{2iT}\times \; &
    \begin{tikzpicture}[baseline=-0.5ex]
            \draw[pblue] plot[domain=0:45, samples=200, smooth, variable=\a] ({0.5*cos(\a)},{0.5*sin(\a)});
            \draw[pgreen] plot[domain=45:135, samples=200, smooth, variable=\a] ({0.5*cos(\a)},{0.5*sin(\a)});
            \draw[pblue] plot[domain=135:180, samples=200, smooth, variable=\a] ({0.5*cos(\a)},{0.5*sin(\a)});
            \draw[pblue] plot[domain=180:360, samples=200, smooth, variable=\a] ({0.5*cos(\a)},{0.5*sin(\a)});
    	 	\draw[pred] (-0.5,0) -- (-0.5-0.5,0);
    	 	\draw[pred] (0.5,0) -- (0.5+0.5,0);
		 	\fill[black] (-0.5,0) circle (0.05);
		 	\fill[black] (+0.5,0) circle (0.05);
            \filldraw[white] (0,0.5) circle (0.1414213562);
            \draw (0,0.5) circle (0.1414213562);
            \draw[black] (-0.1,0.5-0.1) -- (+0.1,0.5+0.1);
            \draw[black] (-0.1,0.5+0.1) -- (+0.1,0.5-0.1);
        \end{tikzpicture} 
    &&= \\ & && \hspace{-4.0cm}  \! \int_{\vec{q}} \frac{\kappa^2 T \gamma_{\phi}^2(p) [ 2\gamma_{\phi}(q)(m^2+Z(q)q^2 + R^{\phi}(q)) + \gamma_{\phi}(r)(m^2+Z(r)r^2 + R^{\phi}(r)) ]  \partial_k R_k^{\phi}(q)}{ (m^2+Z(q)q^2 + R^{\phi}(q))^2 (m^2+Z(r)r^2 + R^{\phi}(r)) [ \gamma_{\phi}(q) (m^2+Z(q)q^2 + R^{\phi}(q)) + \gamma_{\phi}(r)(m^2+Z(r)r^2 + R^{\phi}(r)) ]^2 } \\
    -\frac{i\gamma_{\phi}^2(p)}{2iT}\times \; &
    \begin{tikzpicture}[baseline=-0.5ex]
            \draw[pblue] plot[domain=0:45, samples=200, smooth, variable=\a] ({0.5*cos(\a)},{0.5*sin(\a)});
            \draw[pgreen] plot[domain=45:135, samples=200, smooth, variable=\a] ({0.5*cos(\a)},{0.5*sin(\a)});
            \draw[pblue] plot[domain=135:180, samples=200, smooth, variable=\a] ({0.5*cos(\a)},{0.5*sin(\a)});
            \draw[pblue] plot[domain=180:360, samples=200, smooth, variable=\a] ({0.5*cos(\a)*(1.0-0.1*sin(15*\a))},{0.5*sin(\a)*(1.0-0.1*sin(15*\a))});
    	 	\draw[pred] (-0.5,0) -- (-0.5-0.5,0);
    	 	\draw[pred] (0.5,0) -- (0.5+0.5,0);
		 	\fill[black] (-0.5,0) circle (0.05);
		 	\fill[black] (+0.5,0) circle (0.05);
            \filldraw[white] (0,0.5) circle (0.1414213562);
            \draw (0,0.5) circle (0.1414213562);
            \draw[black] (-0.1,0.5-0.1) -- (+0.1,0.5+0.1);
            \draw[black] (-0.1,0.5+0.1) -- (+0.1,0.5-0.1);
        \end{tikzpicture} 
    &&= \int_{\vec{q}} \frac{g^2 T p^2 q^2 \sin^2\theta\, \gamma_{\phi}^2(q)\, \partial_k R_k^{\phi}(q) }{r^2 \gamma_{j}(r) [ \gamma_{j}(r)/w +  \gamma_{\phi}(q) (m_k^2 + Z_k(q)q^2 + R_k^{\phi}(q) )]^2} \\
    -\frac{i\gamma_{\phi}^2(p)}{4iT}\times \; &
    \hspace{0.5cm} \begin{tikzpicture}[baseline=-0.5ex]
            \draw[pgreen] plot[domain=0:180, samples=200, smooth, variable=\a] ({0.5*cos(\a)},{0.5*sin(\a)});
            \draw[pblue] plot[domain=180:360, samples=200, smooth, variable=\a] ({0.5*cos(\a)},{0.5*sin(\a)});
            \draw[pred] (0,-0.5) -- (-0.5,-0.5);
            \draw[pred] (0,-0.5) -- (+0.5,-0.5);
            \fill[black] (0,-0.5) circle (0.05);
            \filldraw[white] (0,0.5) circle (0.1414213562);
            \draw (0,0.5) circle (0.1414213562);
            \draw[black] (-0.1,0.5-0.1) -- (+0.1,0.5+0.1);
            \draw[black] (-0.1,0.5+0.1) -- (+0.1,0.5-0.1);
        \end{tikzpicture}
    &&= - \int_{\vec{q}} \frac{g^2 T p^2 q^2 \sin^2\theta \, \partial_k R_k^{\phi}(q) }{2(m_k^2 + Z_k(q)q^2 + R_k^{\phi}(q))^2} \bigg( \frac{1}{r^2 \gamma_{j}(r)} + \frac{1}{s^2 \gamma_{j}(s)} \bigg)  \\
    -\frac{i\gamma_{j}^2(p) \mathcal{T}_{lm}(\vec{p})}{2iT(d-1)}\times \; &
    \begin{tikzpicture}[baseline=-0.5ex]
            \draw[pblue] plot[domain=0:45, samples=200, smooth, variable=\a] ({0.5*cos(\a)},{0.5*sin(\a)});
            \draw[pgreen] plot[domain=45:135, samples=200, smooth, variable=\a] ({0.5*cos(\a)},{0.5*sin(\a)});
            \draw[pblue] plot[domain=135:180, samples=200, smooth, variable=\a] ({0.5*cos(\a)},{0.5*sin(\a)});
            \draw[pblue] plot[domain=180:360, samples=200, smooth, variable=\a] ({0.5*cos(\a)},{0.5*sin(\a)});
            \draw[pred,smooth,domain=0:0.5,variable=\x] plot (-\x-0.5,{-0.07*sin(8*3.14159*\x r)});
            \draw[pred,smooth,domain=0:0.5,variable=\x] plot (\x+0.5,{-0.07*sin(8*3.14159*\x r)});
            \node[] at (-0.9,-0.3) {\footnotesize $l$};
            \node[] at (+0.9,-0.35) {\footnotesize $m$};
		 	\fill[black] (-0.5,0) circle (0.05);
		 	\fill[black] (+0.5,0) circle (0.05);
            \filldraw[white] (0,0.5) circle (0.1414213562);
            \draw (0,0.5) circle (0.1414213562);
            \draw[black] (-0.1,0.5-0.1) -- (+0.1,0.5+0.1);
            \draw[black] (-0.1,0.5+0.1) -- (+0.1,0.5-0.1);
        \end{tikzpicture} 
    &&= -\int_{\vec{q}} \frac{g^2 T(p^4+(r^2-q^2)^2 - 2p^2(r^2+q^2)) (\gamma_{\phi}(r) + \gamma_{\phi}(q))^2 \partial_k R^{\phi}_k(q)}{4(d-1) p^2 \gamma_{\phi}(r)[ \gamma_{\phi}(r) (m^2 + Z(r) r^2 + R^{\phi}(r) ) + \gamma_{\phi}(q)( m^2 + Z(q) q^2 + R^{\phi}(q) ) ]^2} \\
    -\frac{i\gamma_{j}^2(p) \mathcal{T}_{lm}(\vec{p})}{4iT(d-1)}\times \; &
    \hspace{0.5cm} \begin{tikzpicture}[baseline=-0.5ex]
            \draw[pgreen] plot[domain=0:180, samples=200, smooth, variable=\a] ({0.5*cos(\a)},{0.5*sin(\a)});
            \draw[pblue] plot[domain=180:360, samples=200, smooth, variable=\a] ({0.5*cos(\a)},{0.5*sin(\a)});
            \draw[pred,smooth,domain=0:0.5,variable=\x] plot (-\x,{-0.5+0.07*(cos(8*3.14159*\x r)-1)});
            \draw[pred,smooth,domain=0:0.5,variable=\x] plot (\x,{-0.5+0.07*(cos(8*3.14159*\x r)-1)});
            \node[] at (-0.4,-0.85) {\footnotesize $l$};
            \node[] at (+0.4,-0.9) {\footnotesize $m$};
            \fill[black] (0,-0.5) circle (0.05);
            \filldraw[white] (0,0.5) circle (0.1414213562);
            \draw (0,0.5) circle (0.1414213562);
            \draw[black] (-0.1,0.5-0.1) -- (+0.1,0.5+0.1);
            \draw[black] (-0.1,0.5+0.1) -- (+0.1,0.5-0.1);
        \end{tikzpicture}
    &&= \int_{\vec{q}} \frac{g^2 T (p^4+(r^2-q^2)^2 - 2p^2 (r^2+q^2)) (\gamma_{\phi}(r) + \gamma_{\phi}(s)) \partial_k R_k^{\phi}(q)}{8(d-1)p^2 \gamma_{\phi}(r) \gamma_{\phi}(s)(m^2 + R^{\phi}(q) + Z(q)q^2)^2} 
\end{alignat*}
with $\vec{r} \equiv \vec{p}-\vec{q}$, and $\vec{s} \equiv \vec{p}+\vec{q}$, and where $\theta$ denotes the angle between $\vec{p}$ and $\vec{q}$.

This constitutes a closed system of flow equations for $\gamma_{\phi,k}(p)$ and $\gamma_{j,k}(p)$, which can be solved on top of the static flow equations discussed above. We choose the UV initial conditions $\gamma_{\phi,\Lambda}(p) = \sigma_{\Lambda}p^2$ and $\gamma_{j,\Lambda}(p) = \bar{\eta}_{\Lambda}p^2$ with $\sigma_{\Lambda}=1$ and $\bar{\eta}_{\Lambda}=1$, and the $k$-independent mode-coupling constant $g=1$.

As stated in the main text, we also consider a variant of our truncation in which we evaluate the flow equations of the kinetic coefficients \eqref{eq:kinCoeffFlow} at the IR minimum of the effective potential. In practice, we first evaluate the flow of the effective potential and read off the IR minimum at $k=0$. In a second step, we then evaluate the flow equations of the kinetic coefficients on this IR minimum. 
We still perform the Taylor expansion of the effective potential \eqref{eq:effPotTaylorExpFiniteH} around the $k$-dependent minimum $\rho_{0,k}$, assuming that the IR minimum is within its radius of convergence. In this regard, we have verified that our result for the scaling function in Fig.~\ref{fig:kawasaki_exp_comp} do not visibly change at the next order in the expansion ($N_{\text{ord}}=6$).
Moreover, we assume a $k$-independent wave function renormalization factor $Z_k(p) = 1$. The reason for this is the following. For consistency, the flow equation \eqref{eq:flowOfZ} of $Z_k(p)$ would also have to be evaluated at the IR minimum. However, in contrast to the kinetic coefficients, $Z_k(p)$ feeds back into the static flow equations. As such, an iterative procedure, for example, would be needed to determine the IR minimum in this case. This is out of scope for the present work, since our main motivation for this second expansion is an estimate for the systematic (truncation) error.

\subsection{Kawasaki approximation}
\label{sec:FRGFlowInKawApprox}

The Kawasaki approximation \cite{Kawasaki:1970dpc} corresponds to a perturbative 1-loop expansion of the real-time FRG flow in the mode-coupling constant $g$. 
In case of the heat conductivity it means that the flow of the shear viscosity and the squared mass is neglected, 
and an Ornstein-Zernike form for static susceptibility is assumed. 
One can verify that in the critical regime the tadpole diagram in Fig.~\ref{fig:flowOf2PtFnc} 
dominates, which to 1-loop order is given by
\begin{equation}
    I_{\phi,k}(p) \equiv - \int \frac{d^dq}{(2\pi)^d} \frac{g^2 T p^2 q^2 \sin^2\theta \, \partial_k R_k^{\phi}(q) }{2(m^2 + q^2 + R_k^{\phi}(q))^2} \bigg( \frac{1}{\bar{\eta} r^4} + \frac{1}{\bar{\eta} s^4} \bigg)   \label{eq:IPhiKawasaki}
\end{equation}
with
\begin{equation}
    r = \sqrt{p^2+q^2-2pq\cos\theta} \quad\text{and}\quad s=\sqrt{p^2+q^2+2pq\cos\theta} \,.
\end{equation}
The contribution $\Delta\gamma_{\phi}(p)$ to the kinetic coefficient of the order parameter is given by integrating \eqref{eq:IPhiKawasaki} from $k=\Lambda$ to $k=0$, which can be performed analytically since the right-hand side of \eqref{eq:IPhiKawasaki} is a total $k$-derivative.  (For simplicity, we take $\Lambda \to\infty$ since the integral turns out to be finite.) Solving the angular integral in $d=3$ spatial dimensions yields
\begin{equation}
    \Delta \gamma_{\phi}(p)= - \frac{g^2 T}{16\pi^2\bar{\eta}p}  \int_0^{\infty} \frac{dq}{m^2+q^2} \bigg( 4pq^2+q(p^2+q^2)\ln\bigg[\bigg(\frac{p-q}{p+q}\bigg)^2\bigg] \,. \label{eq:qIntKA}
\end{equation}
At finite $p\neq 0$, the remaining integral over $q$
can be solved via the residue theorem. To this end, one first extends the lower integration bound to $-\infty$ (which can be remedied by a factor of $1/2$ since the integrand is symmetric in $q$), and, for convenience, one introduces the dimensionless variables $x \equiv p\xi$ and $y \equiv q\xi$,
\begin{equation}
    \Delta \gamma_{\phi}(p)= - \frac{g^2 T \chi_{\phi}}{16\pi^2\bar{\eta} \xi^3} \int_{-\infty}^{\infty} \frac{dy\,y^2}{1+ y^2} \bigg( 2+\frac{x^2+y^2}{2xy}\ln\bigg[\bigg(\frac{x-y}{x+y}\bigg)^2 \bigg] \bigg) \label{eq:zIntKA}
\end{equation}
where we have expressed $1/m = \xi/\chi_{\phi}$ with the correlation length $\xi$ and the static susceptibility $\chi_{\phi}$ of the order parameter.
By closing the integration contour in the upper half complex $y$-plane one picks up a single pole at $y=i$ (cf.~Appendix 6B of Ref.~\cite{Onuki_2002}). Moreover, using $\ln[(x-i)^2/(x+i)^2] = 4i\arctan x$, we find
\begin{equation}
    \Delta \gamma_{\phi}(p)= \frac{g^2 T \chi_{\phi}}{8\pi\bar{\eta} \xi^3} \left( 1+(x-x^{-1})\arctan x \right) \,.
\end{equation}
The singular part of the wavenumber-dependent relaxation rate of the order parameter is thus given by
\begin{align}
    \Gamma^{\phi}(p) &= \gamma_{\phi}(p)(p^2+\xi^{-2}) = \frac{g^2 T}{6\pi \bar{\eta}\xi^3} K(p\xi) 
\end{align}
where we rediscover the Kawasaki function
\begin{align}
    K(x) = \frac{3}{4} \left( 1+x^2+(x^3-x^{-1})\arctan x \right) \,.
\end{align}

\subsection{Critical scaling of the static susceptibility}

In this section, we study the wavenumber ($p$) dependence of the static susceptibility
\begin{equation}
\chi_{\phi}(\tau,h,p) \equiv
    \langle \phi(-\vec{p}) \phi(\vec{p}) \rangle - \langle\phi(-\vec{p})\rangle \langle \phi(\vec{p})\rangle \label{eq:staticSusc}
\end{equation}
which in our truncation \eqref{eq:freeEnergyTrunc} evaluates to
\begin{equation}
    \chi_{\phi}(\tau,h,p) = \frac{1}{Z(\tau,h,p)p^2+m^2(\tau,h)}
\end{equation}
where all quantities are understood at the IR scale $k=0$.

\begin{figure*}[t]
    \centering
    \begin{minipage}{0.34\linewidth}
        \centering
        {(a)}
        \includegraphics[width=\linewidth]{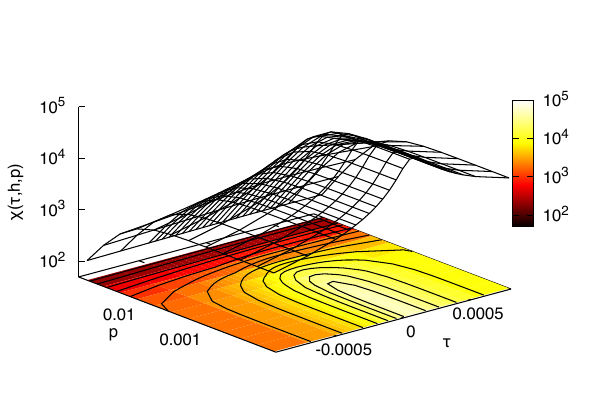}
    \end{minipage}
    \begin{minipage}{0.32\linewidth}
        \centering
        {(b)}
        \includegraphics[width=\linewidth]{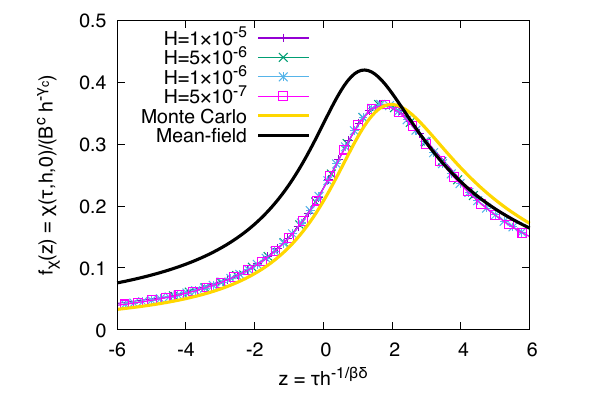}
    \end{minipage}
    \begin{minipage}{0.32\linewidth}
        \centering
        {(c)}
        \includegraphics[width=\linewidth]{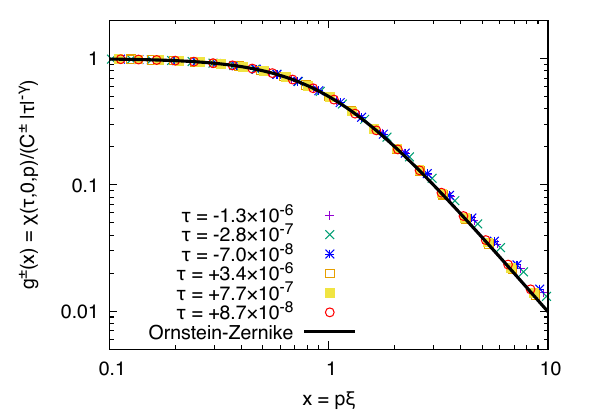}
    \end{minipage}
    \caption{(a) Static susceptibility $\chi(\tau,h,p)$ at fixed $H = 5\times 10^{-7}$ as a function of reduced temperature $\tau$ and wavenumber $p$ within the critical region. (b) Our FRG result for the magnetic-scaling function $f_{\chi}(z)$ of the static susceptibility $\chi(\tau,h,0)$ at $p = 0$, compared to the result of Monte-Carlo simulations \cite{Karsch:2023pga} and the result of the mean-field approximation. (c) Scaling functions $g^{\pm}(x)$ for the universal temperature and wavenumber dependence of the static susceptibility $\chi(\tau,0,p)$ obtained when the critical point is approached for $\tau \to 0^{\pm}$ at $h=0$, in comparison with the Ornstein-Zernike function $g^{\pm}(x) = (1+x^2)^{-1}$. \label{fig:staticSuscScaling}}
\end{figure*}

The magnetic scaling of the static susceptibility at $p=0$ 
follows from another derivative of the order parameter \eqref{eq:phiScaling} with respect to $H$,
\begin{align}
    \chi_{\phi}(\tau,h,0) = \frac{\partial \phi}{\partial H} = B^c h^{-\gamma_c} f_{\chi}(z) \label{eq:chiScaling}
\end{align}
with $\gamma_c = 1-1/\delta$ and 
\begin{equation}
    f_{\chi}(z) = \frac{1}{\delta}\left( f_G(z) - \frac{z}{\beta} f_G'(z) \right) \,. \label{eq:chiScalingFnc}
\end{equation}
The universal function $f_{\chi}(z)$ is normalized according to $f_{\chi}(0) = 1/\delta$. In particular, a comparison of \eqref{eq:chiScaling} at $z=0$ with the expected behavior $\chi = C^c h^{-\gamma_c} + \cdots$ of the susceptibility for $H \to 0^+$ at $T=T_c$ entails that the amplitude ratio $B^c/C^c=\delta$ is universal. This is visible in Fig.~\ref{fig:staticSuscScaling} (b).

The static scaling hypothesis for the temperature and wavenumber dependence of the static susceptibility at $h=0$ reads (with scaling parameter $s>0$)
\begin{equation}
    \chi_{\phi}(\tau,0,p) = s^{2-\eta}\chi_{\phi}(s^{1/\nu}\tau, 0, sp) \,.
\end{equation}
Setting $s^{1/\nu}|\tau| = 1$ yields (with $\gamma = \nu(2-\eta)$)
\begin{equation}
    \chi_{\phi}(\tau,0,p) = |\tau|^{-\gamma}\chi_{\phi}(\sgn\tau, 0, |\tau|^{-\nu} p) \,, 
\end{equation}
which allows us to write
\begin{equation}
    \chi_{\phi}(\tau,0,p) = C^{\pm} |\tau|^{-\gamma} g^{\pm}(p\xi(\tau,0)) \,, \qquad \xi(\tau,0) = f^{\pm} |\tau|^{-\nu} \,,
\end{equation}
where the superscript $\pm$ indicates the symmetric ($+$) and the broken ($-$) phase, respectively. The universal function $g^{\pm}(x)$ depends on the dimensionless variable $x = p\xi(\tau,0)$. The non-universal amplitudes $C^{\pm}$ and $f^{\pm}$ are fixed by requiring the following normalization of the universal functions,
\begin{equation}
    g^{\pm}(0) = 1 \,, \quad \frac{\partial g^{\pm}(x)}{\partial x^2} \bigg\rvert_{x=0} = -1 \,. \label{eq:gNorm}
\end{equation}
In particular, the latter condition means that $\xi$ is identified with the second-moment correlation length \cite{PhysRevB.10.2818}
\begin{align}
    \xi(\tau,h) = \left[-\frac{1}{\chi_{\phi}(\tau,h,p)} \frac{\partial \chi_{\phi}(\tau,h,p)}{\partial p^2} \bigg\rvert_{p=0}\right]^{1/2} \label{eq:2ndMomentCorrLen}
\end{align}
which is a proxy for the `true' correlation length $\xi_{\text{gap}}$ that characterizes the exponential decay of the static susceptibility \eqref{eq:staticSusc} in coordinate space. Near the critical point, the ratio $\xi_{\text{gap}}/\xi$ approaches a universal value. In the $3d$ Ising universality class, it is only slightly larger than one, and the largest value of $\xi_{\text{gap}}/\xi \approx 1.03$ is reached when the critical point is approached from the broken phase at $H=0$ (cf.~Table 11 in Ref.~\cite{Pelissetto:2000ek}).

For the truncation \eqref{eq:freeEnergyTrunc} of the coarse-grained free energy, the correlation length \eqref{eq:2ndMomentCorrLen} evaluates to $\xi =\sqrt{Z(p=0)}/m$. Our FRG results for this quantity are shown in Fig.~\ref{fig:staticSuscScaling} (c), for the two cases of approaching the critical point from the symmetric phase ($\tau\to 0^+$) and the broken phase ($\tau\to 0^-$) at $H=0$. The asymptotic behavior for $x \to \infty$ is given by $g^{\pm}(x) \sim x^{-(2-\eta)}$, in which we find $\eta \approx 0.11$ for the anomalous dimension of the order parameter. For comparison, we have also included the Ornstein-Zernike function $g^{\pm}(x) = (1+x^2)^{-1}$, which appears to describe the overall shape of the FRG quite well, up to small deviations at large $x$ due to the small but finite anomalous dimension $\eta$.

\subsection{Fits for scaling functions}

\subsubsection{Magnetic scaling of thermal diffusivity and shear viscosity}

For the magnetic scaling of the thermal diffusivity \eqref{eq:DpMagneticScaling} shown in Fig.~\ref{fig:Dp_scaling} (b) we use the following fit function which is based on Pad\'e approximants of order $[N_D/N_D]$ in the compactified scaling variable $\arctan z$,
\begin{align}
    f_D(z) =  \bigg( \frac{1+\sum_{j=1}^{N_D} a_j (\arctan z)^j}{1+\sum_{l=1}^{N_D} b_l (\arctan z)^l} \bigg) (1+z^2)^{\nu(\zeta-2)/2} \,. \label{eq:fDFit}
\end{align}
This form guarantees $f_D(0) = 1$, and the limiting behavior $f_D(z) \sim |z|^{\nu(\zeta-2)}$ for $z\to\pm\infty$ (with different universal amplitude ratios for the two limits $z \to \pm\infty$), while remaining an analytic function along the real $z$-axis.
As written in the main text, the magnetic scaling $D_{\phi}(0,h,0) \sim h^{\frac{\nu}{\beta\delta}(\zeta-2)}$ for $h\to 0$ yields $\zeta \approx 3.06$.
Moreover, we derive the critical exponent $\nu \approx 0.643$ from \mbox{(hyper-)scaling} relations with the values of $\beta$ and $\delta$ specified in the main text. Similarly, we use the following fit function for the magnetic scaling  of the shear viscosity \eqref{eq:etaMagneticScaling} shown in Fig.~\ref{fig:eta_scaling} (b),
\begin{align}
    f_{\eta}(z) =  \bigg( \frac{1+\sum_{j=1}^{N_{\eta}} a_j (\arctan z)^j}{1+\sum_{l=1}^{N_{\eta}} b_l (\arctan z)^l} \bigg) (1+z^2)^{-\nu x_{\eta}/2} \label{eq:fEtaFit}
\end{align}
with the critical exponent $x_{\eta} \approx 0.047$ from the main text.
We use $N_D = N_{\eta} =4$ and the fit interval $z\in[-20,20]$, which yields a relative fit error of less than $1\%$ in both cases. The corresponding coefficients $a_1,\ldots,a_4,b_1,\ldots,b_4$ are provided in Table \ref{tab:padeCoeff}.

\begin{table}[b]
\caption{\label{tab:padeCoeff}%
Non-vanishing coefficients for the Pad\'e approximants 
appearing in \eqref{eq:fDFit}, \eqref{eq:fEtaFit}, and \eqref{eq:EPlusFit}.
}
\begin{ruledtabular}
\begin{tabular}{cccc}
&\multicolumn{1}{c}{$f_{D}(z)$}&\multicolumn{1}{c}{$f_{\eta}(z)$}&\multicolumn{1}{c}{$E^{+}(x)$}\\
\colrule
$a_1$ & $-0.536071 $  & $-1.19892  $ & $-1.16471 $ \\
$a_2$ & $-0.328303 $  & $ 0.478348 $ & $ 0.36174 $ \\
$a_3$ & $ 0.102539 $  & $-0.103142 $ &  \\
$a_4$ & $ 0.162053 $  & $ 0.0436332$ &  \\
$b_1$ & $-0.365969 $  & $-1.21508  $ & $-1.16301 $ \\
$b_2$ & $-0.15553  $  & $ 0.480308 $ & $ 0.362261$ \\
$b_3$ & $-0.0100193$  & $-0.0932585$ & \\
$b_4$ & $ 0.20802  $  & $ 0.0385436$ & \end{tabular}
\end{ruledtabular}
\end{table}

\subsubsection{Temperature scaling of wavenumber dependent thermal diffusivity}

In the Kawasaki approximation, the universal dependence of the thermal diffusivity on reduced temperature and wavenumber  is described by the Kawasaki function \eqref{eq:KawFnc}. To fit our FRG results shown in Fig.~\ref{fig:Dp_scaling} (c), we consider the following generalization of the Kawasaki function,
\begin{align}
    K^{\pm}(x) = \frac{3}{4}( b^{\pm} +\frac{4-b^{\pm}}{3} x^2 + (a^{\pm}x^{\zeta} - b^{\pm}/x) \arctan x  ) \label{eq:GenKawFnc}
\end{align}
which ensures the limiting behavior $K^{\pm}(x)/x^2 \to 1$ for $x\to 0$ and $K^{\pm}(x) \sim x^{\zeta}$ for $x \to\infty$. The Kawasaki approximation corresponds to $b=1$, $a=1$ and the dynamic critical exponent $\zeta=3$.

We determine the dynamic critical exponent $\zeta$ from our real-time FRG data by looking for a plateau in the logarithmic derivative $d\log K^{\pm}(x)/d\log x$ for $x \gg 1$, which yields $\zeta \approx 3.06$ in both cases.

Fitting the functional form \eqref{eq:GenKawFnc} to our FRG data for values of $x$ within the scaling region (i.e., for values of $x$ which are smaller than the value of $x$ at which the plateau in the logarithmic derivative stops), we find $a^+=0.582558$, $b^+=-0.276446$ for $\tau \to 0^+$ with a relative fit error of less than $3\%$, and 
$a^-=0.987448$, $b^-=0.639599$ for $\tau\to 0^-$ with a relative fit error of less than $1\%$.

\subsubsection{Temperature scaling of wavenumber dependent shear viscosity}

For the universal dependence of the shear viscosity on reduced temperature and wavenumber shown in Fig.~\ref{fig:eta_scaling} (c) we use the following fit function in the case when the critical point is approached from below ($\tau\to 0^-$),
\begin{align}
    \frac{E^{-}(x)}{x^2} = (1+ax^{nx_{\eta}})^{-1/n} \label{eq:EMinusFit}
\end{align}
which ensures the asymptotic behavior $E^{-}(x)/x^2 \to 1$ for $x \to 0$ and $E^{-}(x) \sim x^{2-x_{\eta}}$ for $x\to\infty$. We determine the critical exponent $x_{\eta}$ of the shear viscosity by looking for a plateau in the logarithmic derivative $d \log E^{-}(x)/d\log x$, which yields $x_{\eta} \approx 0.047$. The best-fit values $a=0.319254$, $n = 36.1146$ yield a function with a relative error of less than $0.1\%$.

For the $\tau\to 0^+$ case we use the following fit function, which uses \eqref{eq:EMinusFit} for the asymptotic behavior for $x\to\infty$ and a Pad\'e approximant of order $[N_E/N_E]$ in the compactified variable $\arctan x$ to describe the non-concavity of the universal function (cf.~Fig.~\ref{fig:eta_scaling} (c)),
\begin{align}
    E^{+}(x) = \bigg( \frac{1+\sum_{j=1}^{N_{E}} a_j (\arctan x)^j}{1+\sum_{l=1}^{N_{E}} b_l (\arctan x)^l} \bigg)E^{-}(x) \,. \label{eq:EPlusFit}
\end{align}
The coefficients shown in Table~\ref{tab:padeCoeff} (with $N_E=2$) correspond to a fit with a relative error of less than $0.1\%$.

\end{document}